# Design and Implementation of User-Friendly and Low-Cost Multiple-Application System for Smart City Using Microcontrollers

**Zain Mumtaz[1], Zeeshan Ilyas[1], Ahmed Sohaib[1], Saleem Ullah[2], Hamza Ahmad Madni[1,*]**

[1] Department of Computer Engineering, Khwaja Fareed University of Engineering & Information Technology,
   Rahim Yar Khan, 64200, Pakistan.
[2] Department of Computer Science, Khwaja Fareed University of Engineering & Information Technology, Rahim
   Yar Khan, 64200, Pakistan.
[*] Corresponding author: 101101770@seu.edu.cn

## ABSTRACT

The need to make the world smarter and safer has led to the growth of intelligent and secure cities that connect the physical world to the virtual world, providing real-time services that improve real-time situations. Here, we proposed the design and experimentally proved a smart city application and products. Our proposed system has seven main contributions, i.e., "Smart street lights", "Smart home", "Bio-metric door and home security system", "Intelligent traffic lights management and road security system", "Private and smart parking", "Intelligent accident management system" and "Smart information display/ notice board system". Our prototypes / products employ Arduino UNO board, Node MCU, Ultrasonic sensor, Fingerprint module, Servo motors, GSM, GPS, LEDs, Flame Sensor, Bluetooth and Wi-Fi module etc. Firstly, we controlled street lights using android application from mobile, that also record data of objects passing on the road. After that, Wi-Fi module through the internet is used for home automation in which the appliances will be controlled with android application. Moreover, fingerprint-based door security system has also been introduced that only operates on the authorized match. Similarly, home security system is proposed in which alarm and window will be automatically opened when fire or suspicious activity detected, while location will also be sent to the security department simultaneously. Furthermore, a smart traffic control system is proposed which adjusts traffic signals during peak hours, skipping capability, and a road security system to detect vehicle verification by OCR technique. Besides, private and smart parking system that only works on authorized detection of vehicles, this system also shows available parking slots. Most importantly, there is an automated accident management system in sensitive areas, that sends the exact location of the incident via text message and automatically throws sprinkles of water on the fire. Finally, a digital notice board system that displays data received from mobile app., climatic and current affairs about the city. We are very confident that our proposed systems of which three are prototype "streetlights control, smart accident management, information display systems" and four are products "home automation, bio-metric security, smart traffic control ,parking systems", are efficient, reliable, and cost-effective and can be easily tested and implemented on a large scale under real conditions, which will be useful in future smart city automation and smart home applications.

## 1. INTRODUCTION

A smart city is an urban environment that utilizes different forms of electronic Internet of Things (IoT) devices to collect data and then use the information obtained from that data to handle properties, energy, and services effectively. This involves data gathered from people, devices, and assets that are processed and analyzed for traffic and transportation systems, power plants, energy management, Networks of water supply, waste control, information systems for crime prevention, schools, hospitals, and other municipal facilities.

The literature survey starts with an analysis of the various descriptions and the connection between of a smart city [1], the relationship between the smart city and IoT [2], while the economic and pricing policies and their links in communication and data collection for IoT [3] is also presented. Surveys [4] on the architecture for smart cities, security aspects of a smart city, and an overview of the smart city deployments [5] around the globe are considered.

In Ref. [6], techniques used by sensors in the Cloud for IoT applications to connect the gap among the Cloud of Things and the Internet of Things. The objective of Fog computing [7] in smart city applications and a view for a smart city application called Electric Vehicles Charging is also discussed. Notable in-depth learning algorithms [8] study presented with video analytics for employment and the utilization of wireless sensor networks in smart cities.

The traditional street light system has only two options of switching: OFF and ON only, which are not effective because this leads to the energy wastage due to continuing to operate on maximum voltage. A lot of electricity wastage when no light needed on the roads, most of the time, streetlights are continuously kept 'ON' due to mechanical problems or by the carelessness of the engineer. In this regard, several automation systems had been proposed to overcome this manual operation. Still, most of them have a lack of performance as it cannot be controlled for specific street lights switching purposes, and the use of IR sensors is deficiency because it cannot work in Sunrays.

A method introduced in Ref. [9] for street lights switching by applying LoRa (Long-range) protocol, which is a typical wireless protocol to use in smart cities due to its low power waste, reliable communications, and long-range indoors and also for outdoors. They used three devices based on the Arduino open-source electronics platform to develop this system; MCDSL (measure and control tool for streetlights), GWLN (gateway LoRa network ), ad an LLMD(lighting level measurement device).

Smart street lighting system is presented in Ref. [10] by using Arduino, LDR, and IR sensors. They introduced DIM lighting as well as high lights on motion detection, which lacks the efficiency due to usage of IR because Sun rays disturb the transfer of Infrared Rays. Also, the street lights cannot be controlled wirelessly for switching on special occasions or specific needs.

A model presented in Ref. [11] that divides street lights energy usage into three categories: low, moderate, and high. The use of street light energy is minimal in the day-time, moderate when regular traffic on roads averages and high when heavy traffic is on roads. Street lights switch "ON" If a vehicle enters in the region after sensing its entry.

Improvement of the energy efficiency and quality of street lighting to obtain power saving in street lights scheme is discussed in Ref. [12], they applied two different solutions: luminous flux regulators installation and luminaires are replaced with LED. Another Method to Monitor and control of street light using GSM [13] technology also proposed, in when if it detects the street light failure, it sends an SMS by using a GSM module to the control unit for informing about the failure. Lights control based on Intensity & Time [14], but the drawback is, during cloudy days, the intensity of light is very low all over the day, and this makes the lamp glow for the entire day, which leads to power loss. Smart and adaptive weather lighting control system with the intelligent embedded system [15], Automatic control system for street light with switch relay [16], and smart street light using wind power is presented [17]. Lighting control system using infrared (IR) obstacle detection sensor [18], light-dependent resistor (LDR) [19], and Arduino microcontroller [20], [21] together are introduced in the past [22], [23]. In the past, the street light automation based on Sun tracking sensors [24], [25], and light-dependent resistors is also used to switch the streetlights on the detection of the sun lights. Moreover, streetlight automation by using solar energy [26] and ZigBee [27] to control streetlights was also implemented. Afar from switching the light ON/OFF, a new approach is proposed to DIM the light (lower than the maximum brightness) [28], [29] during the no traffic hours, which hopefully useful for overcoming the power consumption problem.

Traditional homes are built in such a way that there are no security measurements and an efficient way to interact with things kept inside a house. Such as if you want to turn OFF the fan of some other room when no one is sitting there, you will go to that room and turn that specific OFF. Especially If some aged or disabled person living in the house, how could they be going to operate appliances? So there are many challenges to be faced in traditional homes. As Smart City's vision is evolving, it is getting safer. A secured smart home is made up of technologies for home automation and home security. The technologies for home automation [30] and various household devices are being connected and controlled over the internet [31] is described. Besides using the internet as a platform, integrated home automation systems were built using Bluetooth communications, controlled remotely using android devices [32]. Other platforms include ZigBee communication [33], the Ethernet [34], and the trans-receiver modules[35], which also sometimes used to replace the internet where local transmission is the primary concern. Switching home appliances using Raspberry [36] and Arduino is also discussed. According to the authors in Ref. [37], an essential part of a home security system is to enable various electrical devices and electronic gadgets in it to interact and communicate with each other. Fingerprint door locks are also becoming widely used and could be an excellent alternative to the Wi-Fi [38] or Bluetooth door locks [39]. In Ref. [40], a voice recognition module is used to control a voice-controlled smart home device. An LPG gas leakage and accident safety mechanism has been adopted in Ref. [41]. A multi-level home security framework comprising various sensor nodes, UART (Universal Asynchronous Receiver and Transmitter) and PIC (Priority Interrupt Controller) is proposed [42]. A low-cost home security system captures the data and stores it whenever PIR (Passive InfraRed) sensor senses any human motion is developed in Ref. [43]. A smart locking system using Bluetooth technology and a camera [44], in which movement of the user is captured, and the user will be detected. Then only the user is given a locking or unlocking key for the system.

Traffic management in cities needs careful planning and a traffic control system because there is a massive traffic problem. Roads are limited, vehicles are more, and there are numerous accidents and other kinds of deadly incidents that are caused by inappropriate traffic administration. Traditional traffic signals waste too many people's precious time as well as electricity. Traffic signals for that road keep running, no matter if there is no traffic on the road. Traffic signals timer stays the same for all roads, even if one way has lesser traffic than other routes.

In Ref. [45] summarize a similar data lifecycle analysis for smarter cities within the traffic management systems (TMS) framework. However, it is restrained in its applicability to smart city development, applications, and services since it focused on TMS only. Traffic management data is one of the most significant data sources in a regular smart city where people and the government can benefit significantly from controlling such data and applying a proper analysis [46]. Residents will be able to use traffic data to schedule the time of arrival to a destination [47]. Controlling the traffic lights are based on Density [48], IR sensor [49], an Arduino Uno [50] is proposed, while the Vision-Based technique [51] to monitor traffic also presented. Controlling of traffic using Raspberry-PI [52] and simulation of traffic light [53] system using Arduino and LabVIEW [54] is also performed. In contrast, if someone break signal, an e-mail generation system is also introduced [55]. Apart from modern traffic control systems, the term "internet of things" (IoT) [56] is also presented for controlling traffic lights with the internet that made it feasible to control remotely. Traffic management using the Wi-Fi module [57], PIC microcontroller [58], RF-ID reader [59], GSM module [60] is also introduced while traffic control by digital imaging cameras [61] and license plate recognition [62] for remote monitoring of traffic.

Public parking is a big issue in metropolitan areas in both developed and developing countries. As a result of the increasing incense of car ownership, many cities are dealing with a shortage of car parking areas with a disparity between parking supply and demand, which can be considered a fundamental cause for metropolitan parking problems. These days, finding parking slots in big shopping malls to park the car is too complicated and time-consuming, because there is no such system to display available parking slots.

Vehicles parking based on Wi-Fi technology [63] and the use of Bluetooth Low Energy (BLE) [64] as a protocol to connect sensors and gateways are discussed. Smartphones are also considered in these solutions, mainly to end available spaces [65]. Smart car parking system using Arduino [66], IR sensors [67], Raspberry-PI [68], IoT [69], and with the android mobile application [70] has been proposed. Intelligent parking space detection system based on Image segmentation [71] and parking fee collection based on number plate recognition [72] is also discussed. Smart parking reservation system using RF-ID [73], GSM [74], Ultrasonic [75] based presented while smart parking guidance system using 360 camera and Haar-Cascade Classifier [76] is also introduced. Besides, artificial intelligence [77] to enhance park search but barely identify the technical implementation data.

How did the fire brigade going to know the exact location of that specific area where accidentally fire occurs? In our daily life, we may hear about that some house or factory catches fire due to short circuit, but the team of fire brigade did not reach at the time. There is no such system to inform the police, fire brigade department, or hospital when somewhere accident happens. It is a big issue of communication and knowing the exact location of the area to arrive on time.

Vehicle accident detection system using Arduino [78], Wi-Fi [79], and GPS & GSM [80] in which it gets the location of accident place, and IoT based [81] car accident detection, and notification algorithm for general road accident are proposed. Not only with specialized hardware, accident detection system using image processing [82], MDR [83], machine learning methods such as random forest classifier [84] are also introduced. Using smartphones for detecting car accidents and providing emergency responders with situational awareness [85] and in-vehicle networks through OBD-II devices [86] are discussed.

As a workload increases, people do not have time to watch television or to read newspapers for any information about the city, such as news, events or weather information, etc. Giving awareness to the people is a susceptible task that needs some specific ways to make them educated about what happens in the city and needs some display boards only for the people of the town. Peoples become unaware of daily incidents in their town because there is no such system to let people know about their city. The news channel and social media display worldwide news that might be irrelevant for them, as they want to know the current affairs in their town, such as load shedding schedule for a specific area.

In literature, the electronics notice board using Arduino [87], Raspberry-Pi [88], Bluetooth [89], and design of an information display based on several LED [90] is proposed.

Development of a speech [91] and SMS controlled dot-matrix smart noticing system with RF transceiver [92] module is presented. All the described system works in short-range, so to overcome this issue, displaying of information with ZigBee [93], Wi-Fi [94], IoT [95] and GSM [96] has been introduced.

As far as we know, a need still exists for the design of smart city systems that support various embedded applications system such as street lights automation, homes automation/security, traffic lights management, private and smart parking, information display, and accident management system, which is very easy to use and can be easily assembled in modest hardware circuits. In this paper, we proposed the design and experimentally demonstrated a smart city application and products; the proposed system have seven main contributions "Smart street lights," "Smart home" "Bio-metric door and home security system," "Intelligent traffic lights management and road security system," "Private and smart parking," "Intelligent accident management system" and "Smart information display/ notice board system". Firstly, we propose and experimentally demonstrate a design to construct a street lights automation system based on a click on the cellphone with an Android operating system. In the proposed automation scheme, the street lights can be controlled with the mobile application to turn ON/OFF all or specific lights for particular purposes, such as on events. This work is accomplished with the proper arrangements of the microcontroller Arduino Uno, Bluetooth, real-time clock. Also, it will count the number of vehicles passing through the road and switching of lights, which will be shown live on the serial monitor of Arduino IDE [20], local website, and also stored in an excel file. Thus, the proposed android mobile-based system is designed and illustrated, utilizing a lab-scale model to prove that the proposed devices can be easily implemented on large-scale soon.

Next, we propose and experimentally demonstrate a design to construct an IoT based automation system for home to control the appliances while sitting anywhere in the world with just on a click on the cellphone with an Android operating system using the internet. Moreover, home security and a biometric door system are also installed to prevent suspicious activity. This work is accomplished with the proper arrangements of the Node MCU, relay module, ultrasonic sensor, smoke sensor, biometric module and android application.

Next, we propose and experimentally illustrated design to construct an intelligent traffic lights control system which having capability of skipping the street lights if there is no traffic detected on the road and turn OFF all traffic signals during No traffic hours. Moreover, a traffic security system is also installed to prevent suspicious activity/non-registered vehicles. This work is accomplished with the proper arrangements of the Arduino, timer display, ultrasonic sensor, camera and OCR library (Emgu CV).

Next, we propose and experimentally illustrated design to construct an RF-ID based private parking system in which only authorize person will park the car if he/she has the unique key. Moreover, a smart parking system which works on vehicle detection and also has an available/parked LED indicator and display available parking slots on the LCD. This work is accomplished with the proper arrangements of the Arduino, counter display, ultrasonic sensor, and RF-ID.

Next, we propose and experimentally illustrated design to construct an accident management system, if it detects the fire, it will start sprinkle of water and send the exact position via SMS to the fire brigade station. Moreover, three special buttons are installed on the road, if the incident happens, press the reverent button, and Police station/ fire brigade station or hospital will be notified with the location. This work is accomplished with the proper arrangements of the Arduino, GSM module, GPS module, smoke sensor and water pump.

Finally, we propose and experimentally illustrated design to construct a smart information display system which can show the daily news/ information or notices in a city correctly received wirelessly form an android mobile application. Moreover, City live temperature and humidity will also be shown on display to let people more knowledge about the current city situation. This work is accomplished with the proper arrangements of the Arduino, LCD, temperature and humidity sensor and Bluetooth control mobile application.

Besides, Our system will also count the number of times appliance toggle/ door open and suspicious activity, the number of time traffic signals toggles and complete record of vehicles passed through road, the number of times parking slots used/ private car park, and the number of time which buttons pressed and how much accidents happen which will be shown live on the serial monitor of Arduino IDE, website and also stored in excel file. Thus, all proposed systems are designed and illustrated using lab-scale models to show that the proposed products can be implemented in large-scale in the near future.

The innovative element of the proposed work in this paper is Arduino that is a user-friendly microcontroller and can be readily available on demands. At the other hand, automatic systems may be built with Raspberry-PI, ZigBee, and other microcontrollers that are expensive and difficult for the encapsulation method to integrate the different functions in a basic hardware circuit. Also, the purpose of this work is to make life easier for people of old age or physically disabled who are unable to walk and faced with difficulties in carrying out their everyday activities such as controlling home appliances. Thus, this proposed research aims to construct a multiple-functional smart city with a low-cost that benefits peoples in their daily lives.

For the simplicity of analysis, Fig. 1 demonstrates the overview of the proposed smart city having a smart building, fire control, accident management, traffic management, smart display, while smart street lights and smart parking that is connected through sensors and wireless networks.

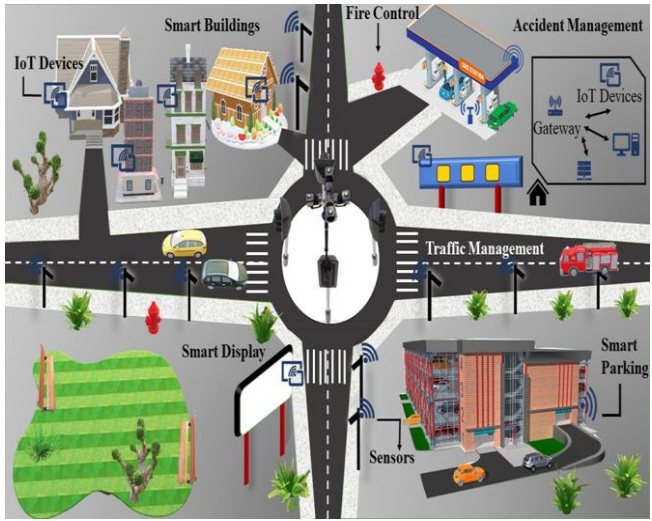

**FIGURE 1.** Simple schematic of low-cost smart city.

This paper is organized as follows. Section 2 provides the idea of the low-cost smart city with a detailed explanation of the electronic components that are used in the proposed system. Section 3 discusses the circuit diagrams and the experimental results of the proposed systems in a lab-scale prototype. Section 4 concludes the paper. Section 5 presents future work.

## 2. MATERIALS AND METHODS

Fig. 2 demonstrates the complete ordering process and the features of the proposed smart city, In Fig. 2a, the streetlights can be controlled with Bluetooth mobile application in which the switching of streetlights only depends on the signal received from the application. All the street lights will be switched OFF if the signal received from mobile application is "All OFF', or unique light switching if the corresponding button is pressed. Arduino also counts the total number of LEDs glow and objects that cross the street in the night, with the help of an ultrasonic sensor, and saves in excel files and demonstrates this to the serial monitor and website.

In Fig. 2b, the switching of appliances with the mobile application it is represented. In this, the user will press the button from Wi-Fi based mobile application, and the signal will send wirelessly to the node MCU. After receiving the signal, it will judge and send a command to the relay module to turn ON the corresponding appliance. Node MCU also counts the total number of appliances switching and saves in excel files and demonstrates to the serial monitor and website.

Fig. 2c illustrates the biometric door and home security system in which the door will only open on valid fingerprint and also detects suspicious activity. First, the biometric module gets the fingerprint of user and matches with predefined fingerprints; if the id seems valid, then the Arduino sends a signal to the motor module to turn ON/OFF the door. If the security system is ON, and the ultrasonic sensor detects some motion within its range, then the Arduino will send a signal to turn on the alarm. Smoke detection sensor senses the value to smoke and sends to Arduino. Upon receiving the data, Arduino translates it to different discrete values from 0–1023 and decides if the obtained value is above the threshold point (a maximum value that is set separately by the consumer from the range of discrete values: 0–1023); it will then be considered as fire, so the alarm and window will be opened. Arduino also counts the total number of the door opened/closed, suspicious activity and saves in excel files and demonstrates this to the serial monitor and website.

Fig. 2d shows that the intelligent traffic control system, in which if the ultrasonic sensor cannot detects any vehicle at the road, then Arduino will send a signal to skip the corresponding road signals. If there is no traffic detected on any of the streets, then no traffic signal will be work. When any vehicle passed form the traffic signal, the camera will capture the image of the number plate. After processing and applying different OCR techniques, it will get the numbers (characters) of the vehicle, and it will compare with its database to check the record of the vehicle. It the legal record found, it will save it, and if no record/ criminal record found, it will turn on the alarm. Arduino also counts the total number of traffic signal switched and saves in excel files and demonstrates this to the serial monitor and website.

Fig. 3a shows that the complete architecture of a parking system having two modes; private parking in which the door will only open if it detects valid ID. The second mode is smart parking in which the LEDs and LCD will indicate the availability and presence of vehicles. In private parking, the user will scan its card to the RF-Id card reader, after scanning value, it will send to Arduino and compared with predefined values. If the id matches with pre-stored ids, the door will be opened, and LCD will display the corresponding parking slot. In smart parking system, if the ultrasonic sensor detects no vehicle in its range, then corresponding Red LED will glow and LCD will display available parking slots, or if the ultrasonic sensor detects the vehicle, the Green LED will shine, and parking slots value will be minus. If four ultrasonic detects no vehicle, the LCD will display four value from total available parking slots. Arduino also counts the total number of private and smart parking slots used and saves in excel files and demonstrates this to the serial monitor and website.

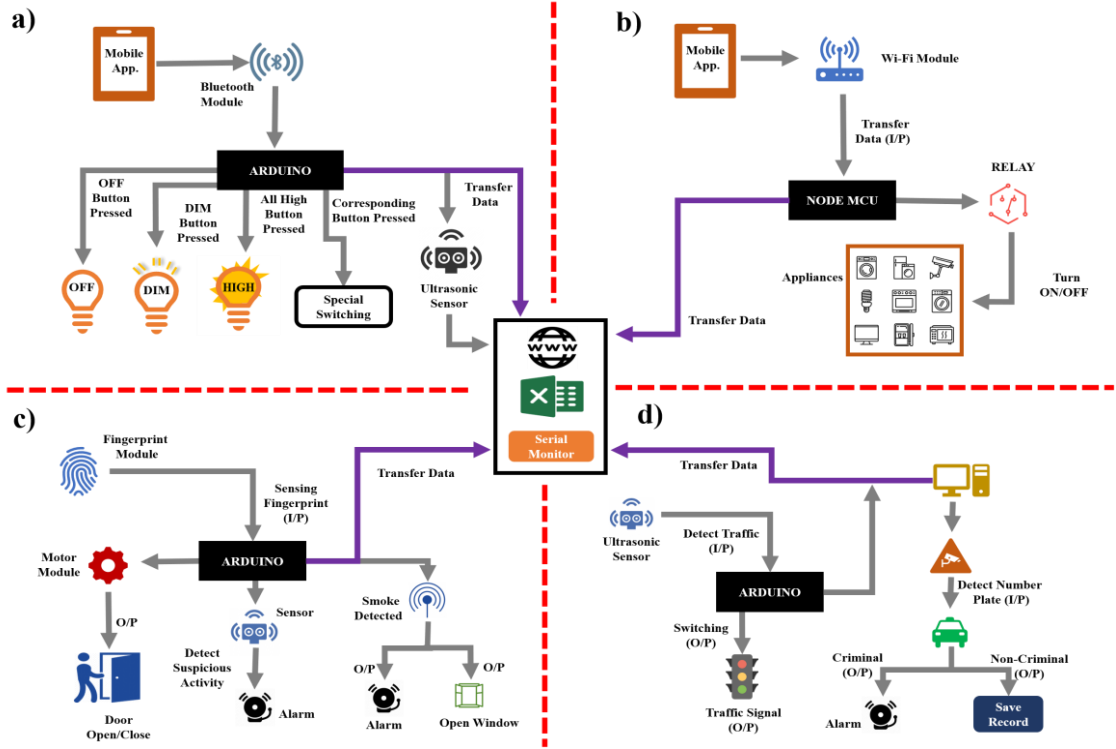

**FIGURE 2.** The architectural design of smart city; (a) Shows the architecture of automatic street lights; (b) Display the architecture design of home automation; (c) Shows the architecture of fingerprint door and security system; (d) Display the architecture design of the traffic control system.

In Fig. 3b, if the fire detection sensor detects the presences of fire, it will send a signal to Arduino to turn ON the water pump and alarm. Arduino will also get the accurate location of the incident through the GPS module and send to the fire brigade station through SMS. Moreover, there is three-button (police, ambulance and fire brigade) is installed on the road, if the user pressed any of the buttons, then the Arduino will get GPS location where the button pressed and send through SMS to the relevant department. Arduino also counts the total number of incident happened, and button pressed and saved in excel files and demonstrates this to the serial monitor and website.

In Fig. 3c, the temperature and humidity sensor sense and send value to the Arduino. After receiving value, it converts it in standard (Celsius and Percentage) and send this to LCD for display. Another mode is to display data received from the android application; first, the user will write information or data that wants to show and press the send button on application through Bluetooth. Arduino will receive data from Bluetooth and send it to LCD for displaying. Arduino also counts the value of temperature and humidity at the current time, data received from the mobile application and saves in excel files and demonstrates this to the serial monitor and website

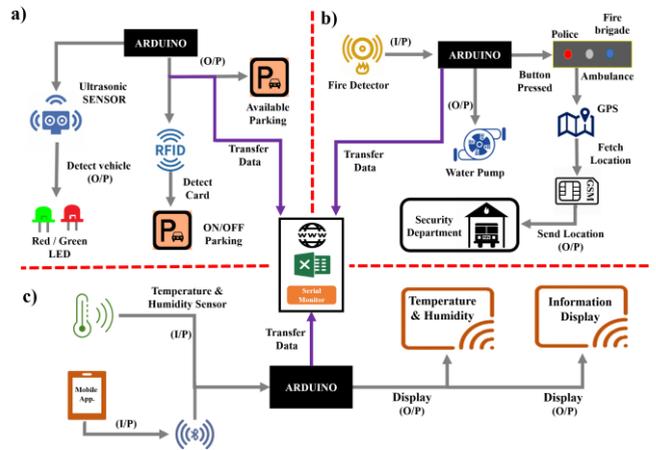

**FIGURE 3.** The architectural design of smart city; (a) Shows the architecture of the smart and private parking system; (b) Display the architecture of smart accident management; (c) Shows the architecture design of smart information display.

### 2.1. ELECTRONIC COMPONENTS

Different electronic components are used for the construction of electronic circuits. Consequently, our proposed circuit diagrams also include the components listed in Table 1.

## 2.1.1. ARDUINO UNO

The Arduino Uno microcontroller board [20, 21] are typically based on the ATmega328 series of microcontrollers and has a desktop and web IDE (integrated development environment) for writing, compiling and uploading programming language codes to memory. Different sensors transfer the data as input to the microcontroller and send feedback to various devices such as motors, LEDs, relay units. It comprises a total of 28 pins from which 14 digital input/output pins (six are PWM pins) and six analogue pins used for contact with electronic components such as LDR sensor, ultrasonic sensors, three pins for grounding and other pins for 5V, 3.3V, AREF (analogue reference), RESET and VIN. Arduino microcontroller has 32 KB of main memory, 2 KB of static random-access memory (SRAM) storage and only 1 KB of read-only electrically erasable programmable read-only memory (EEPROM). Arduino microcontroller has 32 KB of ROM ram, 2 KB of static random access memory (SRAM) and only 1 KB of electrically erasable read-only programmable memory (EEPROM). Arduino primarily supports C / C++ programming language compiler (supports other languages such as Python, Java by libraries), macro-assemblers, and test kits. It also has a USB interface jack for computer control, an external power input port, 16 MHz ceramic resonators and an ICSP (in-circuit serial programmer) header, a reset button to switch to the factory settings. Its operating voltage is 7 to 12V with a maximum of up to 20V.

TABLE I
SPECIFICATION OF ELECTRONIC COMPONENTS USED IN TO DESIGN THE PROPOSED SYSTEM

| Components | Specifications |
|---|---|
| Arduino UNO [20,21] | 28 pins; Operating voltage: 7–12V |
| Node MCU [97] | 17 pins; Operating voltage: 7–12V |
| LCD display [98] | 4 pins; 16*2 display |
| Temp. and Humidity sensor [99] | 4 pins; Accuracy ±1%. |
| Rf-ID reader [100] | 8 pins; Operating voltage: 3.3V, Range: 5cm |
| Bluetooth module HC-05 [101] | 6 pins; Operating voltage: 3.3–5V; Transmission range: 100 m |
| Servo motor [102] | Operating voltage: 5V; Max power: 25W |
| Ultrasonic sensor [103] | Operating voltage: 5V; Range: 4m; Angle: 15º |
| GSM module [104] | 12 pins; Operating voltage: 4V; |
| GPS module [105] | Operating voltage: 3–5V; Baud Rate: 9600 |
| Finger Print Sensor [106] | 8 pins; Operating voltage: 3.6–6V; Fingerprint: 1024 |
| LDR [107] | Operating voltage: 5 V; Range 2–30 cm; Angle: 35º |
| Relay module[108] | Pins: 6; Operating voltage: 5V |
| Android mobile application [109] –[111] | Android compatible |

## 2.1.2. NODE MCU

Node MCU [97] is an open-source IoT framework, with a kernel that operates on the Espressif Non-OS SDK on the ESP8266, and hardware based on the ESP-12 board. The device has 4 MB of flash memory, 64 KB of Sram, 80MHz of machine clock, roughly 50k of available RAM and a Wi-Fi Transceiver chip. It is based on the project eLua which is designed on the ESP8266 Espressif Non-OS SDK. Its voltage operation is 3.3V, while Input Voltage is 7–12V and built-in Wi-Fi; IEEE 802.11 b/g/n. It has total 17 GPIO pins (General Purpose Input/Output), four GND pins, three 3.3V pins, one analogue pin, two reserved pins, and MOSI, CS, MISO, SCLK, EN, RST and Vin pin.

## 2.1.3. LCD DISPLAY

The LCD [98] is an electronic display module that uses liquid crystal to create a visual image. The I2C 16x2 LCD translates a display of 16 characters per line in 2 such lines and very simple module widely used for DIYs and circuits. It has a total of four pins; VCC, GND, SLC and SDA pin.

## 2.1.4. TEMP AND HUMIDITY SENSOR

The DHT11 [99] is a widely used sensor for temperature and humidity. The sensor works with a designated NTC for temperature determination and an 8-bit microcontroller for processing of temperature and humidity values as serial data. The sensor is easy to communicate with other microcontrollers as well as Arduino. The sensor can calculate the temperature from 0°C to 50°C and humidity from 20% to 90% with a precision of ±1°C and ±1%. It has a total of four pins; VCC, Data, N/c and GND.

## 2.1.5. RF-ID

RC522 RFID [100] module is based on NXp's MFRC522 IC. It usually comes with an RFID card chip and a 1 KB plastic fob key chip. The RC522 RFID Reader module is programmed to generate a 13.56MHz electromagnetic field that is used to interact with RFID tags (ISO 14443A standardized tags). The reader can interact with the microcontroller using a 4-pin Serial Peripheral Interface (SPI) with a maximum data rate of 10Mbps. It also facilitates contact through I2C and UART protocols. Its Operating Supply Voltage is 2.5 V to 3.3 V with Max Operating Current 13–26mA and Read Range of 5cm. It has a total of eight pins; SDA, SCK, MOSI, IRQ, GND, RST and VCC.

## 2.1.6. BLUETOOTH MODULE HC-05

The HC-05 [101] Bluetooth module is designed for personal wireless serial connectivity and used in a master or slave configuration, providing it with an excellent solution for wireless communication. This serial port Bluetooth module is fully adequate Bluetooth V2.0 + EDR 3 Mbps modulation with 2.4 GHz radio transceiver and baseband. It contains total six pins; ENABLE pin to toggle within AT and Data command mode, VCC pin for giving voltage, ground pin, TX-Transmitter and RX-receiver for sending and receiving serial data and lastly, a state pin for checking of Bluetooth pairing/un-pairing). Its operating voltage is 3.3–5V and the transmitting range is up to 100 m.

### 2.1.7. SERVO MOTOR

A Servo motor [102] is an electrical device which can push or rotate an object with high precision. If the user wants to turn an item at a certain angle or distance, use the servo motor. It is just a small motor that operates via a servo system. Operating voltage is 3.3V –5V typically; torque is 2.5kg/cm, operating speed is 0.1s/60° and rotation angle is 0°–180°. It has three wires; brown for grounding, red for VCC and orange for PWM signal is given in through this wire to drive the motor.

### 2.1.8. ULTRASONIC SENSOR

An ultrasonic sensor [103] is a tool that utilizes ultrasonic sound waves to measure the distance to an object. An ultrasonic sensor utilizes an amplifier to transmit and receive ultrasonic pulses that convey information about the location of the target. High-frequency sound waves bounce from the border to create distinct echo patterns. Four pins need to communicate with the sensor; VCC, Trig (signal output pin), Echo (signal input pin) and GND. Working voltage is 5V with a max range of 4m, and measuring angle is 15 degree.

### 2.1.9. GSM MODULE

SIM800L [104] is a mini-cellular module that allows GPRS to transmit, send and receive SMS and make and receive voice calls. Low cost and compact footprint and quad-band frequency help make this package the ideal solution for any project involving long-range communication. After connecting the power panel, automatically log in and scan for a cellular network. Onboard LED shows contact status (no network coverage-fast blinking, logging in-slow blinking). Be equipped to tackle a huge power demand of up to 2A. The nominal UART voltage in this module is 2.8V. Higher voltage is going to destroy the machine. This package with two antennas, first one is made with wire (which welds directly to the NET pin on the PCB), very handy in narrow areas. Second PCB antenna with double-sided tape and a pigtail cord with an IPX connector. Its supply voltage is 3.8V – 4.2V, and Working temperature range is -40° to + 85° C. It has total 12 pins; Net, VCC, RST, RXD, TXD, GND, Ring, DTR, MIC+, MIC-, SPK+ and SPK.

### 2.1.10. GPS MODULE

The NEO-6MV2 [105] is a GPS (Global Positioning System) device used for navigating purposes. The device tests its position on earth and provides output data which is the longitude and latitude of its location. These compact and cost-effective receivers provide a wide variety of networking options in a small (16 x 12.2 x 2.4 mm) box. The lightweight architecture, power and memory choices make NEO-6 modules suitable for battery-operated portable devices with very tight cost and space constraints. Its revolutionary architecture gives NEO-6MV2 excellent navigation efficiency, even in the most demanding environment. Its power supply range is 3V to 5V with the default baud rate of 9600 bps. It has a total of four pins; VCC, RX, TX and GND.

### 2.1.11. FINGERPRINT SENSOR

R305 [106] is a fingerprint sensor device with a TTL UART interface. The user can save the fingerprint information in the device and customize it in 1:1 or 1:N mode to identify the individual. The fingerprint device will communicate directly with a 3v3 or 5V microcontroller. For PC interfaces, a level converter (like MAX232) is required. Its operating power is 3.6V–6.0V and stores up to 1024 fingerprint ids. It has a total of eight pins; VCC, GND, TD, RD, VCC, D-, D+ and GND.

### 2.1.12. LDR

The resistance of the LDR [107] depends on the intensity of sunlight impinging on it, and the resistance provided by the sensor drops with a rise in light strength and raises with a decrease in light intensity. LDR is used to measure day-time and night-time, and when sunlight falls on it, it will be called day-time, and when no sunlight falls on it, it will be called night-time. These are very beneficial, especially in light/dark sensor circuits, which help in automatically switching the street lights (ON/OFF).

### 2.1.13. DC RELAY MODULE

Relay Driver [108] is a programmable logic controller is used to manage solid or mechanical state relays in DC and AC voltage power systems. It mainly works as a switch for electronics for ON and OFF. It has six pins; VCC, GND, Input pin, normally open, normally closed and common pin.

### 2.1.14. ANDROID MOBILE APPLICATION

An Android mobile application is a software developed in a computer programming language (C, C++, Java, etc.) which run on the Android platform. The application for controlling the streetlights [109], Home automation [110] and for Smart information display [111] is available and can be easily downloadable.

## 3. DESIGNING METHODOLOGY

Fig. 4a shows the circuit design of automatic street light control system based on Android mobile application using Arduino Uno having a feature of DIM light capability. In this scenario, the lights will switch according to the button pressed on mobile. In this task, an LDR sensor, Sixteen LEDs, eight Ultrasonic sensors, Bluetooth and a single Arduino Mega have been used. One leg of the LDR sensor is attached to the Arduino analogue PIN A0, and the other end to the 5V and the same is attached to the GND port of Arduino. Also, the baseline value for LDR is set to 15 of the discrete values (0–1023) to decide if it is day or night-time. After that, all the positive terminals of the LEDs' set are connected to PINs D2, D3, D4, D4, D6, D7, D8 and D9 as the outputs of the Arduino signals. In this regard, one set of LEDs consists of two independent LEDs. Moreover, connected the GND of all the LEDs' to GND port as illustrated in Fig. 4a. The Trig terminals (represented by green lines) of ultrasonic sensors are attached to the Arduino port from PIN D23, D25, D27, D29, D31, D33, D35 and D37, respectively, and the echo terminals (represented by Brown lines) are connected to the Arduino port from pin number D22, D24, D26, D28, D30, D32, D34 and D36, respectively, which is the input signal to the Arduino board. Likewise, the GND of all ultrasonic sensors is linked to the GND port, and all ultrasonic sensor VCCs are linked to the 5V pin of Arduino. Tx and Rx pin of Bluetooth module are connected to Arduino port number D10, D11 and VCC to 5v and Ground to GND port of Arduino.

Fig. 4b shows the circuit design of the Smart home control system that can be controlled with internet based on Android mobile application using Node MCU.

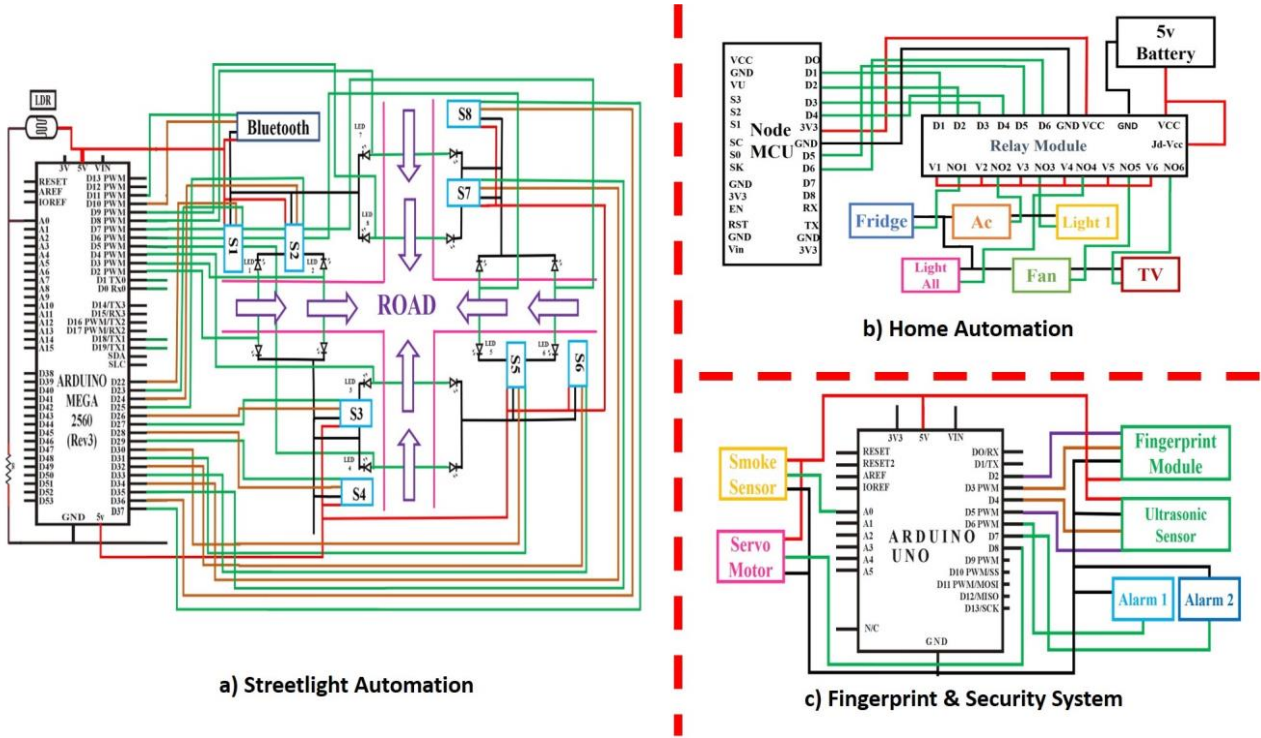

**FIGURE 4.** Circuit diagram of smart city; (a) Shows the schematic of automatic street lights system; (b) Display the schematic design of home automation; (c) Shows the schematic of fingerprint door and security system.

In this scenario, the appliances will turn ON only on the signal received from mobile; otherwise, lights will remain OFF. In this task, a fridge, AC, lights, fan, TV, six-channel relay module and a single Node MCU have been used. D1–D6, VCC and GND Pin of relay module are connected to D1–D6, VCC and GND port to the Node MCU. VCC, JD-VCC and V1–V6 pins of relay module to Positive terminal of 5V battery and GND pin to the negative terminal of the battery. Furthermore, Positive terminal of Fridge, AC, Light1, Light2, fan and TV to NO1–NO6 (normally open) pins of relay module and negative terminal to the negative terminal of the battery as shown in the Fig. 4b.

Fig. 4c shows the circuit design of smart home security and fingerprint door lock system based using Arduino Uno. In this scenario, the alarm will turn on if it senses suspicious activity and door will only open on the valid fingerprint. In this task, a fingerprint sensor, one ultrasonic sensor, two-alarm, one smoke detection sensor, one servo motor, and a single Arduino UNO has been used. TX pin (purple color), RX pin (Brown color), VCC, and GND pin of fingerprint module is connected to D2, D3, VCC, and GND port of Arduino Uno respectively. Furthermore, trig pin (purple color), echo pin (brown color), VCC, and GND pin of the ultrasonic sensor is connected to D5, D4, VCC, and GND port to Arduino Uno as shown in the Fig. 4c. The positive terminals of Alarm1, 2 are connected to D7, D8, and negative terminal to the GND port of Arduino Uno. The green terminal of smoke sensor and servo motor is connected to A0 and D8 port of Arduino Uno, VCC, and GND is connected to 5V and GND port of Arduino Uno.

Fig. 5a shows the circuit design of an intelligent traffic control system using Arduino Mega. In this scenario, the system will skip that traffic signal if no traffic detected on the corresponding road. In this task, four timer module, four ultrasonic sensors, twelve LEDs, and a single Arduino Mega have been used. CLK pin (Purple color) of timer module 1–5 is connected to D2, D4, D6, and D8, respectively, DIO pin (Brown color) to D3, D5, D7, and D9 of Arduino Mega. Moreover, VCC and GND pin of timer module 1–4 is connected to 5V and GND port to Arduino Mega. In the same way, trig pin (blue color) of ultrasonic sensor 1–5 is connected to D14, D16, D18, and D10, respectively, echo pin (green color) to D15, D17, D19 and D11 of Arduino Mega. Moreover, VCC and GND pin of ultrasonic sensors 1–5 are connected to 5V and GND port to Arduino Mega. Furthermore, the Positive terminal of LED 1–12 is connected to D49– D38, respectively, as shown in Fig. 4b.

Fig. 5b shows the circuit design of private and smart parking system using Arduino Uno and RF-ID. In this scenario, the LCD will display available parking slots, and private parking will only open if a valid card is shown. In this task, a five ultrasonic sensor, two servos, one four-digit display, one RF-ID card reader, and a single Arduino mega have been used. Trig pin (purple color) of ultrasonic sensors 1–5 is connected to D11, D9, D7, D5, and D3, respectively. Echo pin (brown color) of ultrasonic sensors 1–5 to D10, D8, D6, D4 and D2 port of Arduino Mega. In the same way, VCC and GND pin of ultrasonic sensors 1–5 are connected to 5V and GND port of Arduino Mega. Furthermore, CLK (purple color), DIO (brown color), VCC, and GND pin of 4 digit display to D31, D30, 5V, and GND port to Arduino Mega as shown in the Fig. 5d. Moreover, Green, VCC, and GND terminal of the servo motor is connected to D43, 5V, and GND port of Arduino Mega. At lastly, VCC, SCK, MOSI, MISO, RST, SDA and GND pin of RF-ID module is connected to 5V, D38–DD42, GND port of Arduino Mega.

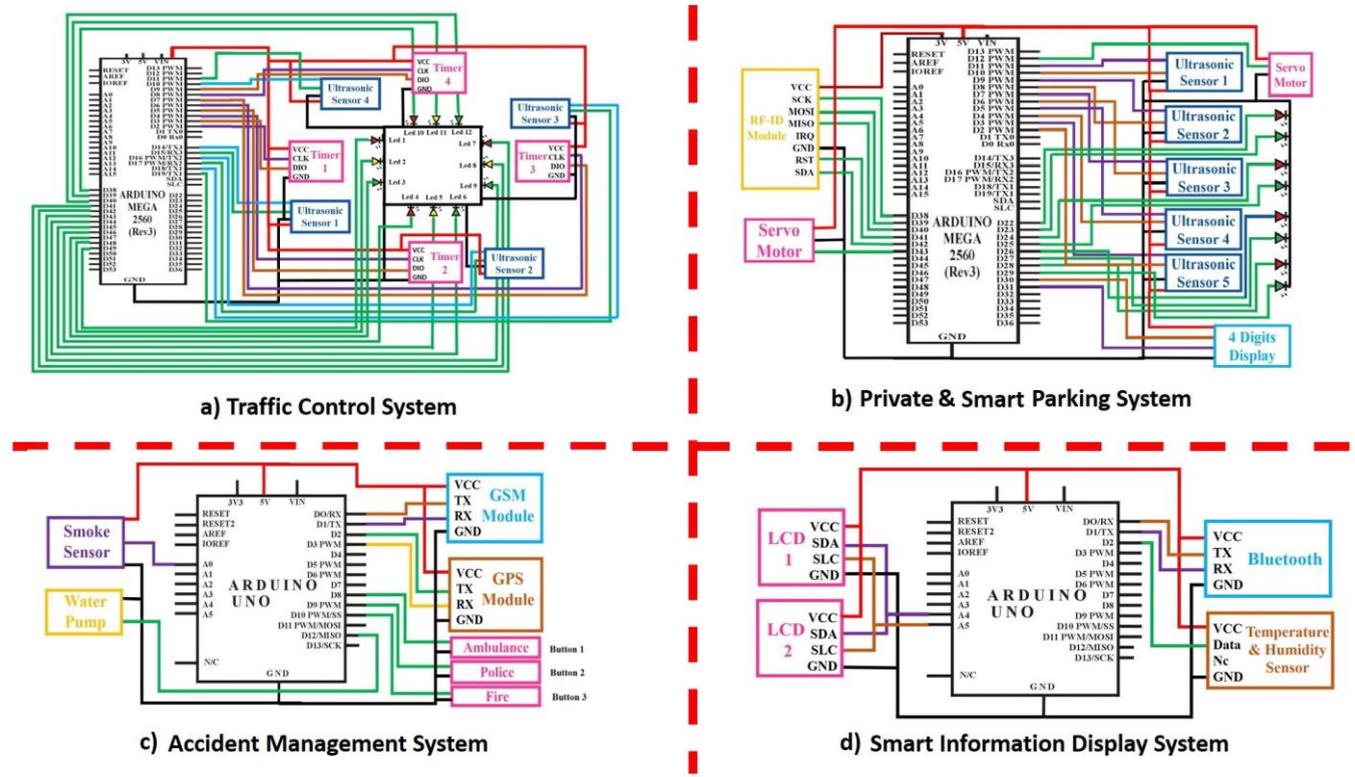

**FIGURE 5.** Circuit diagram of a low-cost smart city; (a) Display the schematic design of the traffic control system; (b) Shows the schematic of the smart and private parking system; (c) Shows the schematic of smart accident management; (d) Display the schematic design of smart information display system.

Fig. 5c shows the circuit design of an intelligent accident management system using Arduino Uno. In this scenario, the system will send accurate location if fire detected or relevant button pressed. In this task, one GSM module, one GPS module, three push-buttons, one smoke sensor, one water pump, and a single Arduino Uno has been used. RX pin (purple color), TX pin (brown color), VCC, and GND pin of the GSM module is connected to D1, D0, 5V and GND port of Arduino Uno. Similarly, RX pin (yellow color), TX pin (green color), VCC, and GND pin of GPS module is connected to D3, D2, 5V and GND port of Arduino Uno.

Moreover, the positive terminal of button 1–3 is connected to D8, D9, D10 as the input signal, and negative terminal to the GND port of Arduino Uno. In the same way, the positive terminal of the water pump to D12, and the negative terminal is connected to the GND port of Arduino. Lastly, data pin (purple color), VCC, and GND pin of the smoke sensor are connected to Analogue A0, 5V, and GND port of Arduino Uno, as shown in Fig. 5c.

Fig. 5d shows the circuit design of the smart Information/notice display system using Arduino Uno. In this scenario, the LCD will display data only when received from the android mobile application. In this task, one Bluetooth module, One Dht11 temperature, and humidity sensor, two 16x2 LCDs, and a single Arduino Uno have been used. TX pin (brown color), RX pin (purple color), VCC and GND pin of Bluetooth module is connected to D0, D1 5V and GND port to Arduino Uno. Data pin (green color), VCC, and GND pin of temperature and humidity sensor are connected to D2, 5V, and GND port of Arduino Uno. In the same way, SDA (purple color), SLC (brown color), VCC, and GND pin of LCD 1–2 is connected to Analogue Pin A4, A5, 5V, and GND port of Arduino as shown in the Fig. 5d. The detailed program coding for each case is provided in supplementary materials.

## 3.1. ANDROID MOBILE APPLICATION

MIT App Inventor [112] is a visual programming drag and drops platform for designing and development of fully functional android mobile applications. App Inventor's user interface is consists of two parts: a designer to choose the components of the application and a blocks-editor for setting the operations and working for the application. App. Inventor's building blocks are simple user interface contains elements such as buttons, labels, list pickers, images, etc., linked with the mobile device's features (Bluetooth, texting, NFC, GPS, etc.) Therefore, the fundamental structures of this drag and drop enabled app developers to efficiently manage the functionalities of these portable, touch-enabled sensing devices. By concentrating on the device's services, app. inventor presents an automatic programming metaphor. A texting component is used for an application that sends and receives texts. The block for identifying an incoming text is "Texting.MessageReceived". This understandable, action-based, drag and drop, event-driven programming model reduces the difficulty level that usually experienced in traditional text-based programming environments. In our application, we have used Bluetooth client component, notifier component, text to speech component, button component, label, and title components.

Fig. 6a shows the main designer view of the MIT application development platform includes the user interface menu, viewer, components menu, and properties menu. In Fig. 6b, the blocks for all the components are shown, such as in screen 1, Bluetooth components blocks having Bluetooth connection, show alert notifier, label component, and speak message components, as shown in Fig. 6b i. In Fig. 6b ii, the blocks for the speech recognition page is shown having a speech recognition component, send text, and speak message components. In Fig. 6b iii, the blocks for pattern one buttons are illustrated having sent text component and label component, which will send "I" when the button is pressed and set a label to forwarding. The blocks for the UP button are shown in Fig. 6b iv, in which the send text component and label component are used, which will send "A" if the user pressed the button and set a label to forwarding.

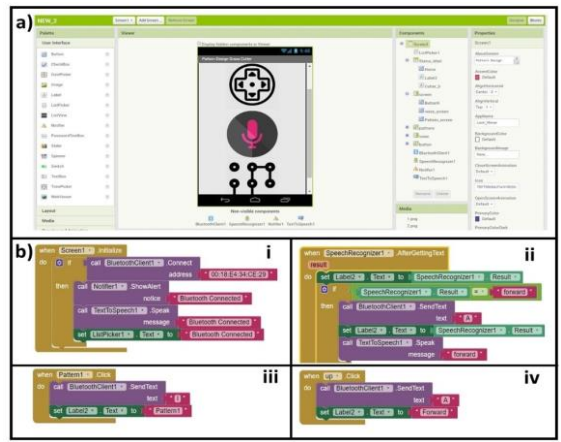

**FIGURE 6.** Design view and blocks editor of android application for pattern design grass cutter; (a) Design view window; (b) i. Bluetooth module components; ii. Speech recognition component; iii. Send text component for the pattern; iv. Send text component for up arrow button.

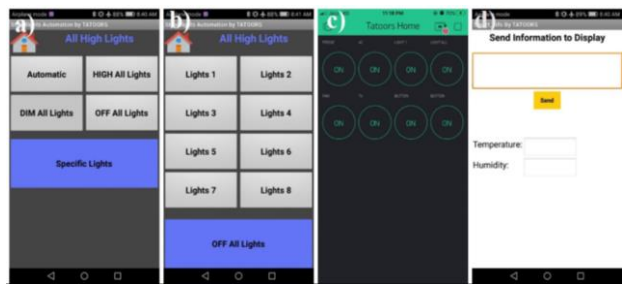

**FIGURE 7.** User Interface of android mobile applications; (a) The homepage of an android application showing five different options to control streetlights; (b) Eight different buttons for specific lights; (c) The homepage of an android application showing eight other options to control home appliances; (d) The homepage for the smart display.

Fig. 7 shows the user interface of the android mobile application where Fig. 7a is the home screen of the smart street light system having six buttons (home, automatic, HIGH all, DIM all, OFF all and specific lights) and one label which display the text when the corresponding button pressed. Fig. 7b showed the interface when the user pressed the specific lights button and eight different buttons to control street light. If the user wants to turn ON particular lights, he/she will push the corresponding button from the application, and also a turn OFF all button to switch OFF all lights. Fig. 7c shows the user interface of the smart home application having eight different buttons from which six is currently used for the appliances such as TV, fridge, AC, etc. Fig. 7d displays the user interface of smart information display having a text editor in which the information/ message is input and a send button that sends the data.

## 4. RESULTS AND DISCUSSIONS

### 4.1. ANDROID MOBILE APPLICATION BASED STREETLIGHT CONTROL SYSTEM

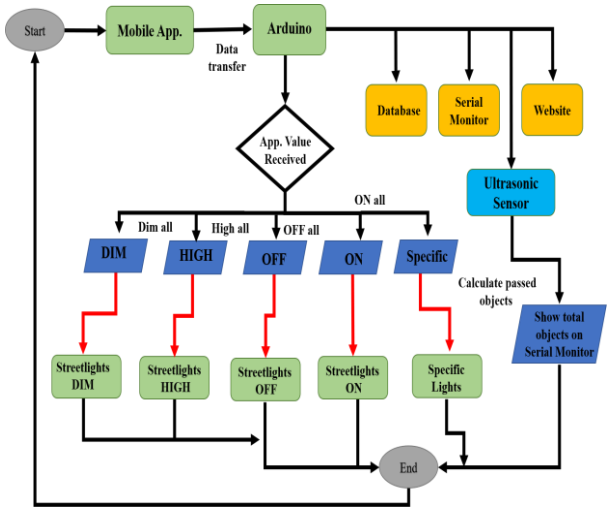

**FIGURE 8.** The flow diagram of the street lights automation system.

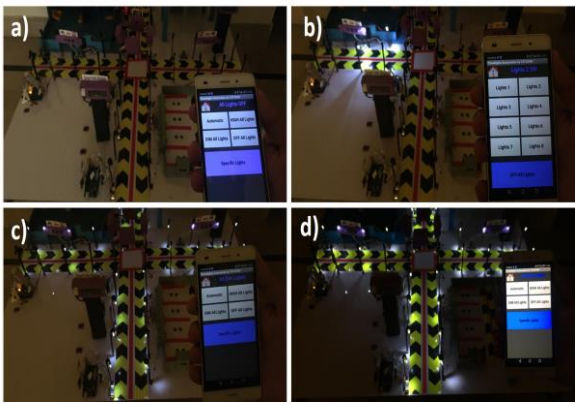

**FIGURE 9.** Result diagrams of the android mobile application based streetlight control system; (a) The user pressed All lights OFF button, so the LEDs are not glowing; (b) The user pressed the Lights 2 button, so the corresponding HIGH LEDs are glowing; (c) DIM LEDs are glowing because the user pressed All DIM button; (d) All HIGH LEDs are glowing as the user pressed HIGH All button on android application. The complete demonstration video of the proposed system is available online [113].

Fig. 8 shows the flow diagram to control the street lights with an Android-based mobile application. In this mode, after establishing the connection to the Android mobile application with Arduino, via the Bluetooth module, sends the corresponding signal to Arduino Uno anytime the user clicks either of the buttons in the program. After the signal is received, Arduino will check this with predefined instruction for turn ON all lights, turn OFF all lights, DIM All lights, or turn ON specific lights. Furthermore, Arduino will count the total number of vehicles passed through the road and the number of time operations done by the android application and save it in the database and will show in the serial monitor of Arduino and website.

Fig. 9 shows the final demonstration of the proposed streetlights system that can be controlled with an Android mobile application using Bluetooth and Arduino Mega. Fig. 9a represents that no lights are glowing because the user pressed ALL Lights OFF button on the android application. Fig. 9b shows the second lights are shining to HIGH, as the user pressed LIGHT 2 button from the mobile app. In Fig. 9c, All lights are in DIM position, because the user pressed ALL lights DIM on the mobile application. Moreover, when the user pressed "ALL lights High" on the android application; as a result, all lights are glowing in a HIGH position, as seen in Fig. 9d. These results demonstrate the efficiency of the proposed idea and give immediate validation for the proposed model. This streetlight system is a prototype and this kinds of application can be implemented in hotels, malls and homes.

### 4.2. SMART HOME USING WI-FI

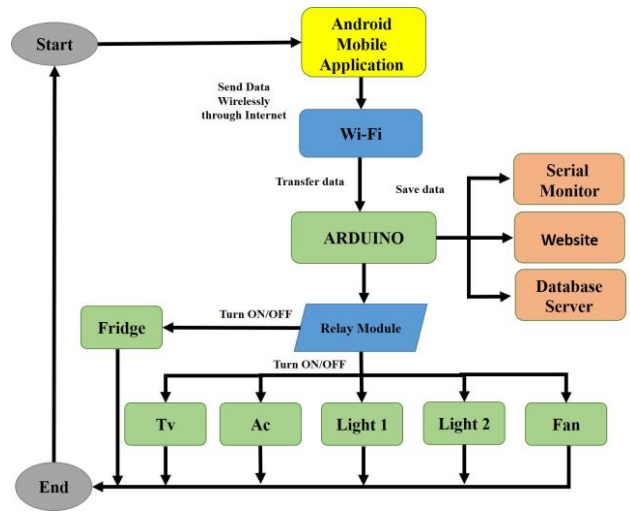

**FIGURE 10.** The Flow Diagram of IoT based smart home.

Fig. 10 shows the flow diagram of a smart home using Wi-Fi and the internet. After establishing the connection to the Android mobile application with Node MCU. Whenever the user clicks either of the buttons in the application, the resulting signal will be sent to the Node MCU. After receiving the signal, Arduino will check this with predefined instructions for the fridge, AC, TV, light 1, lights 2, and fan, then send the command to the relay module to turn ON/OFF the corresponding appliances. Furthermore, Arduino will count the total number of Appliances switching and the number of time operations done by the android application and save it in the database and will show in the serial monitor of Arduino and website.

Fig. 11 shows the final demonstration of the proposed smart home automation system that can be controlled with an android mobile application using Wi-Fi and the Internet. Fig. 11a represents that none of any appliances is working because the user nothing pressed on the android application. Fig. 11b shows the fridge is turned ON, as the user pressed fridge button from the mobile button. In Fig. 11c, AC is in working position, because the user pressed the AC button on the mobile application. Moreover, When the user pressed the LCD and fan button on the android application; as a result, fan and LCD are working as seen in Fig. 11d. Fig. 11e represents that light number 1 and 2 are working because the user pressed the relevant button on the android application. Fig. 11f shows all appliances are turned ON, as the user pressed all buttons on the android application. These results demonstrate the efficiency of the proposed idea and give immediate validation for the proposed model. This home automation system is a product and can be implemented in hotels, malls and homes.

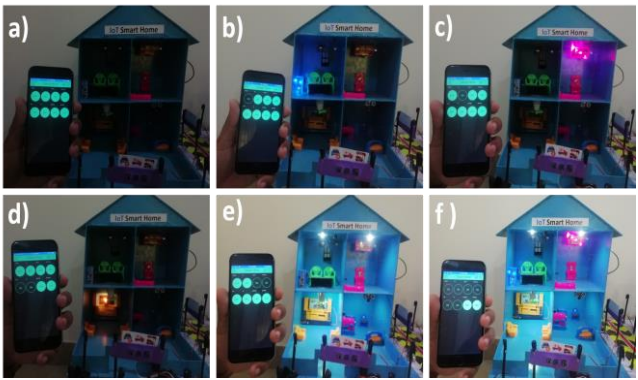

**FIGURE 11.** Result diagrams of the android mobile application based home automation system; (a) The user pressed nothing, so appliances are turned OFF; (b) The user pressed the fridge button, so the fridge is turned ON; (c) AC is working because the user pressed the AC button; (d) Fan and LCD are turned ON as the user pressed fan and LCD button; (e) Light 1 and Lights 2 is working because the user pressed the corresponding button; (f) The user pressed all buttons, so all appliances are Turned ON. The complete demonstration video of the proposed system is available online [113].

### 4.3. BIOMETRIC DOOR AND INTELLIGENT SECURITY SYSTEM

Fig. 12 shows the flow diagram of bio-metric based door and intelligent security system. In the bio-metric door system, when the user scans its fingerprint to open the door, this fingerprint will first go the Arduino Uno, and it will be compared with predefined fingerprints (a total of 1024 Fingerprint can be saved). If the fingerprint matches with saved id's, Arduino will send a signal to the motor module to turn ON the door (the door will be closed automatically after 5 seconds of its opening). In the security system, when the user went out of the home and turned ON the security system, if the ultrasonic sensor detects suspicious activity within its range (10 cm), it will turn ON the alarm. Similarly, when the smoke detection sensor detects the smoke, it will turn ON the alarm and also opens the window. Furthermore, Arduino will count the total number of times the door opened, suspicious activity, and smoke detected and saved it in the database and will show in the serial monitor of Arduino and website.

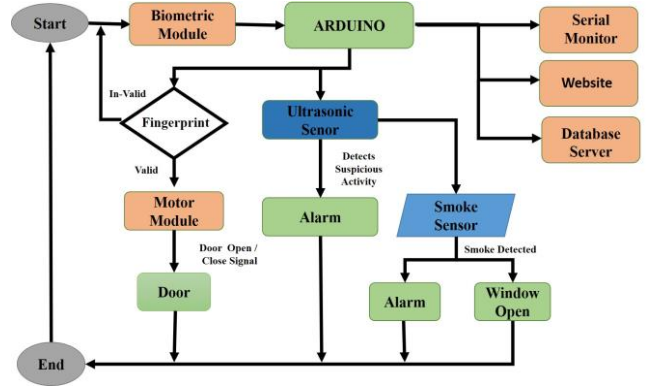

**FIGURE 12.** The flow diagram of the biometric door and smart security system.

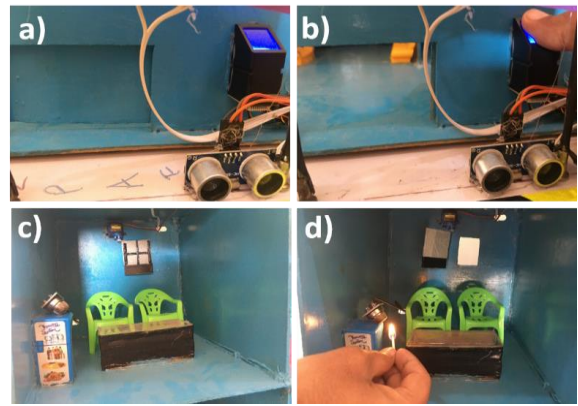

**FIGURE 13.** Result diagrams of the intelligent home security system and biometric door system; (a) The door is closed as no fingerprint detected; (b) The door is opened because the valid fingerprint is detected; (c) The Window is not opened because no smoke detected; (d) Alarm and window are opened as smoke detected. The complete demonstration video of the proposed system is available online [113].

Fig. 13 shows the final demonstration of the proposed intelligent home security and biometric door lock system. Fig. 13a represents that door is in the closed state as the fingerprint module not detected or it is waiting for valid fingerprint. Fig. 13b shows the biometric module detects correct fingerprint, so the door is opened (closed after 5 seconds of opening). In Fig. 13c, alarm and window are closed, because no smoke or fire detected by the smoke sensor. Moreover, flame detected by a smoke sensor; as a result, window and alarm are in the working position as seen in Fig. 13d. These results demonstrate the efficiency of the proposed idea and give immediate validation for the proposed model. This fingerprint door system is a product and this kinds of application can be implemented at homes, buildings, offices, etc.

### *4.4. SMART TRAFFIC CONTROL AND ROAD SECURITY SYSTEM*

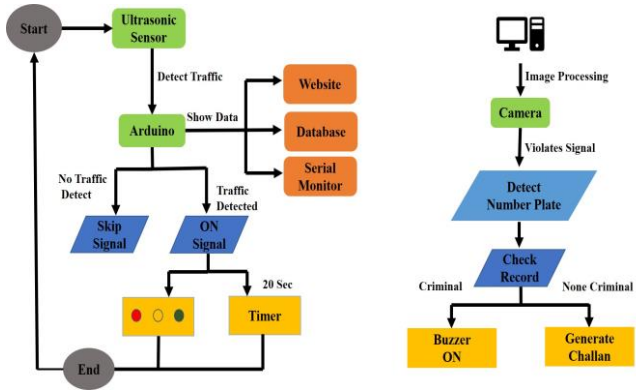

**FIGURE 14.** The flow diagram of the smart traffic control and road security system.

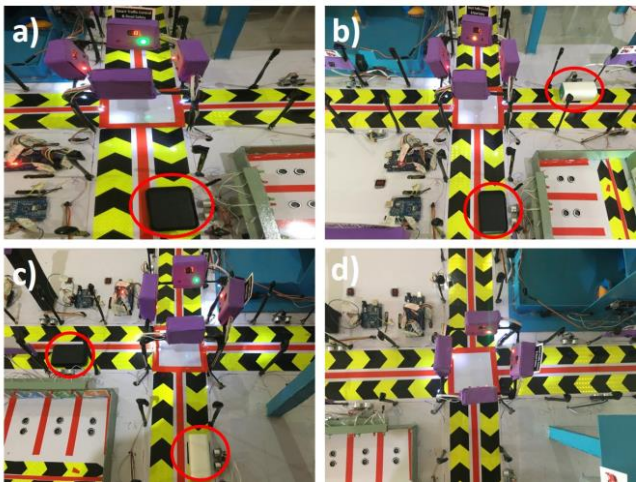

**FIGURE 15.** Result diagrams of the intelligent traffic control system; (a) The sensor detects a vehicle, so the signal is working; (b) The sensor detects a signal on two roads, so the previous signal is in switching position; (c) Next road signal is working because the sensor detects traffic on the road; (d) No traffic signal is working as no vehicle detected. The complete demonstration video of the proposed system is available online [113].

Fig. 14 shows the flow diagram of an intelligent traffic control system that can skip the signal if no traffic detected on the road and also detect un-registered vehicles using the OCR technique. In the traffic control system, when the ultrasonic sensor sense any traffic on the road, it will turn ON that corresponding traffic signal for 20 seconds. Similarly, if the ultrasonic sensor detects no traffic on the road, it will skip that corresponding road signal. In clear terms, all traffic signals will work if traffic detected on all routes, or if there is traffic only on one road, that road signal will operate in the loop. Furthermore, Arduino will count the total number of times the door opened, suspicious activity, and smoke detected and saved it in the database and will show in the serial monitor of Arduino and website. In a traffic security system, it will check all vehicles passing through the road by using the OCR technique by detecting number plates. When a vehicle passed through road, it read the number plate and check this in the database record. It will turn ON the alarm if no data found in the database of that vehicle.

Fig. 15 shows the final demonstration of the proposed intelligent traffic management system that can skip the road signal if no traffic detected on the road. Fig. 15a represents that the ultrasonic sensor detects the vehicle on the road, so the corresponding road signal is working (it will remain in the loop if no traffic is seen on others road). Fig. 15b shows that the ultrasonic sensor detects traffic on two streets, so the previous road signal is in switching position. In Fig. 15c, the next road signal is working now, because the relevant ultrasonic sensor detects traffic on the road (these two road signals will work in the loop if no traffic noticed on another street). Moreover, no traffic detected on any of the streets; as a result, no traffic signal is in the working position, as seen in Fig. 15d. These results demonstrate the efficiency of the proposed idea and give immediate validation for the proposed model. This system is a product and this kinds of application can be implemented at the roads of the city and town.

### *4.5. INTELLIGENT VEHICLE LICENSE PLATE VERIFICATION*

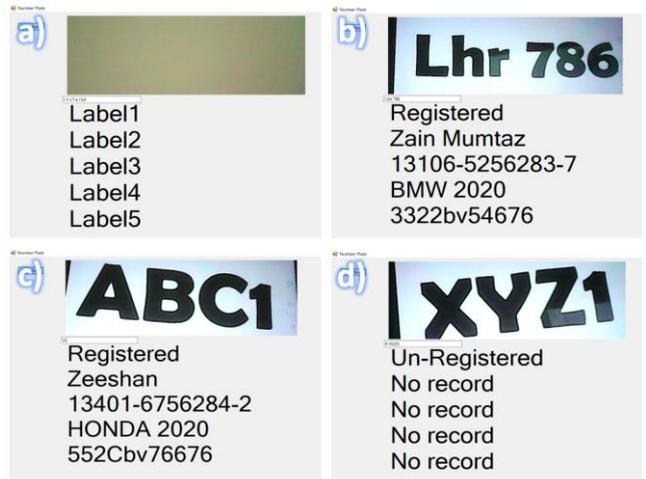

**FIGURE 16.** The Working diagrams of the Vehicle verification system using OCR techinque. The complete demonstration video of the proposed system is available online [113].

Fig. 16 shows the final demonstration of the intelligent vehicle license plate verification using OCR technique. Fig. 16a represents that no license plate detected on the camera, so it is not shows nothing. Fig. 16b shows that the system detected "Lhr 786" license plate, so, it shows the registered data according to it. In Fig. 16c, registered data is showing in the software, because the system detects the "ABC1" license plate. Moreover, an un-registered license plate is detected; as a result, un-registered alert is showing at the software, as seen in Fig. 16d. This vehicle verification technique can be directly implemented in road cameras.

### 4.6. SMART PARKING SYSTEM

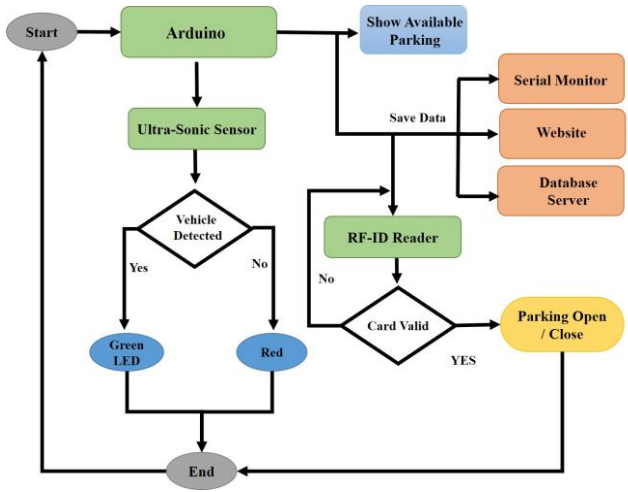

**FIGURE 17.** The flow diagram of the private and smart parking system.

Fig. 17 shows the flow diagram of RF-ID based parking in which parking slots will only open if the valid card is detected and smart parking in which the system displays the remaining parking slots and by LEDs. In a private parking system, the user will show the card to use private parking; RF-Id reader will read the card and send it to Arduino. After receiving the card, it will check its value into predefined values, if the value matches in predefined values, then the parking gate will be open, or if the car mismatches, then the parking will not be opened for the user. In smart parking, there are two types of lights; green light when parking slot is not free and red light when the parking slot is free. In the beginning, if the ultrasonic sensor detects that there is a vehicle on the parking slot, Arduino will send the signal to turn ON the corresponding greenlight and also display available parking slots. If the ultrasonic sensor detects no vehicle on the parking slot, the red light will turn ON, and this system works the same for all parking slots. Furthermore, Arduino will count the total number of times particular parking slots and private parking slots used and save it in the database and will show in the serial monitor of Arduino and website.

Fig. 18 shows the final demonstration of the proposed private and smart parking system that has the capability to show available parking slots. In Fig. 18a, the private parking gate is closed, and LCDs "Show your card", because the RF-ID is waiting for a valid card. Moreover, a valid card is detected by the RF-ID reader; as a result, the private parking gate is opened, and LCDs the corresponding parking spot number as seen in Fig. 18b. Fig. 18c represents that ultrasonic sensor detects the vehicle on the gate, so the parking gate is opened (closed after 5 seconds) and LCD showing four available parking lots because of no vehicle detected by an ultrasonic sensor in parking spots. Fig. 18d shows the gate is closed, and LCD is showing two available parking spots because ultrasonic sensors detected two vehicles on the parking spots. These results demonstrate the efficiency of the proposed idea and give immediate validation for the proposed model. These systems are products and this kinds of application can be implemented in the parking of home, schools, shopping mall, etc.

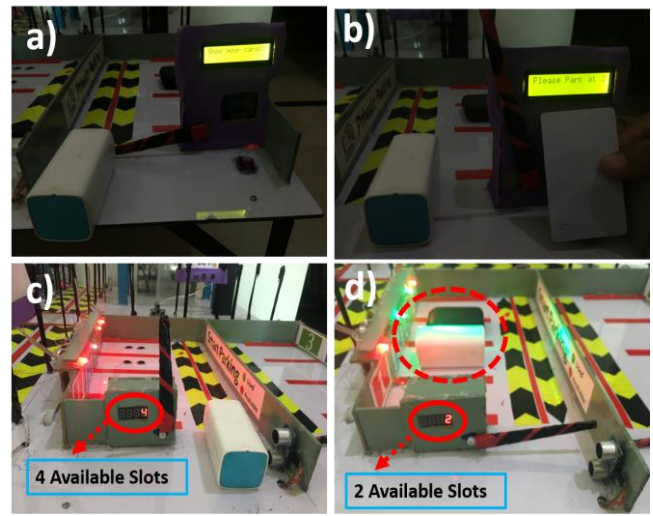

**FIGURE 18.** Result diagrams of the private and smart parking system; (a) Parking gate is not opened because it is waiting for a valid card; (b) The door is opened as the RF-ID reader detects the valid card. (c) The sensor detects a vehicle, so the gate is opened; (d) LCD showing two available parking slots, as two vehicles used parking; The complete demonstration video of the proposed system is available online [113].

### 4.7. SMART ACCIDENT MANAGEMENT SYSTEM

Fig. 19 shows the flow diagram of the smart accident management system that can send an accurate location of the incident through SMS. The system has two modes; automatic and manual. In an automated system, the fire detection system detects a fire; it will send value (0–1023 depending on the sensitivity of fire) to the Arduino. After receiving the signal, it will send turn ON signal to the water pump, alarm, and also it sends an accurate GPS location through SMS using GSM module to the fire brigade station. In the manual mode, there is a specific button installed on the road that can be pushed to send coordinates of that location, such as the police button, ambulance button, and fire brigade button. For example, if some groups of peoples start fighting on the road, the other can push the police button, and the exact location of this incident will be sent to the police within seconds. Furthermore, Arduino will also count the number of times a particular accident happens and save it in the database and will show in the serial monitor of Arduino and website.

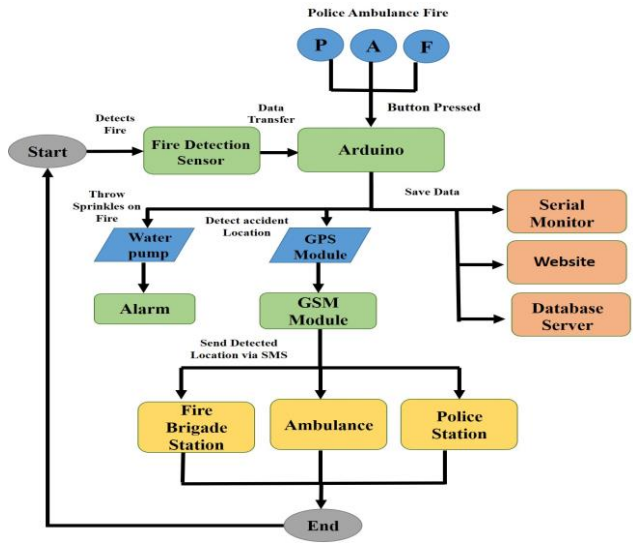

**FIGURE 19.** The Flow Diagram of the Smart Accident Management System.

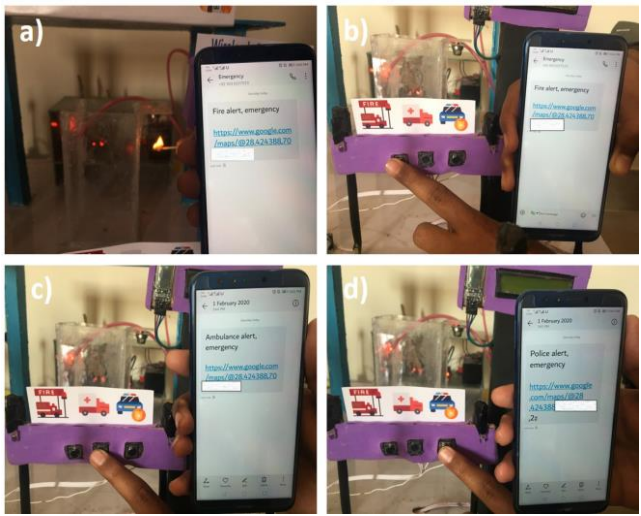

**FIGURE 20.** Result diagrams of the intelligent accident management system; (a) The sensor detects a fire, so the water pump and alarm is working, and the current location is sent; (b) Fire detected location is sent, as the fire button is pressed; (c) Ambulance alert and location is sent because the relevant button is pressed; (d) The police alert and location is received as police button is pressed. The complete demonstration video of the proposed system is available online [113].

Fig. 20 shows the final demonstration of the proposed intelligent accident management system that can send an accurate location on the incident. Fig. 20a represents that the flame sensor detects the fire, so the alarm and water pump are working, and exact GPS location is sent to the fire brigade department through SMS. Fig. 20b shows the exact GPS location and fire alert SMS is sent to fire brigade because the fire button is pressed. In Fig. 20c, the exact GPS location and ambulance alert SMS are sent to the hospital because the ambulance button is pressed. Moreover, the Police button is pressed; as a result, the exact GPS location via SMS is sent to the police station, as seen in Fig. 20d. These results demonstrate the efficiency of the proposed idea and give immediate validation for the proposed model. This system is a prototype and can be rendered in the homes, schools, and streets of towns and communities.

### 4.8. SMART INFORMATION DISPLAY SYSTEM

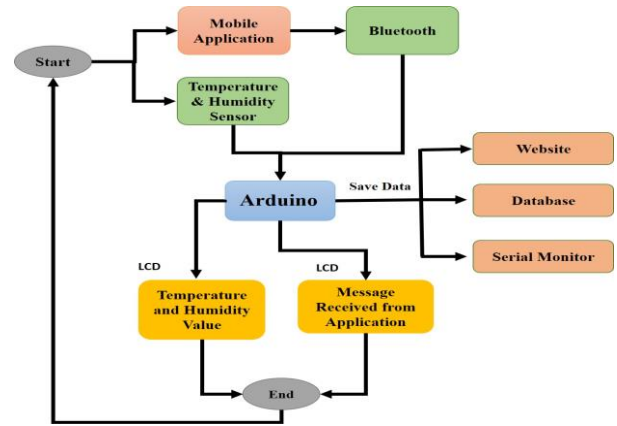

**FIGURE 21.** The flow diagram of the smart information/notice display system.

Fig. 21 shows the flow diagram of the smart Information displaying system in which the current atmosphere and data received from a mobile application is displayed on the LCDs. In the beginning, the temperature and humidity sensor will send data to the Arduino; after receiving data, it will show this on the LCD screen. Similarly, the Bluetooth module will send data through a mobile application to the Arduino. After receiving the data, the Arduino will display the data directly on the LCDs. Furthermore, Arduino will also get temperature and humidity values (10-second interval) and data received from the mobile application and save it in the database and will show in the serial monitor of Arduino and website.

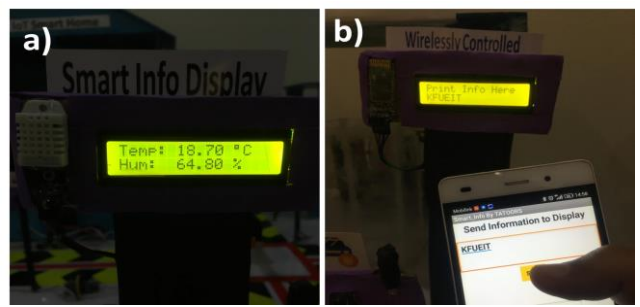

**FIGURE 22.** Result diagrams of the smart information display system; (a) LCD temperature and humidity data; (b) LCD showing received information from the android mobile application. The complete demonstration video of the proposed system is available online [113].

Fig. 22 shows the final demonstration of the proposed smart information displaying/ notice board system that can show wirelessly received data. Fig. 22a represents that LCD is displaying current temperature and humidity value (refreshes after 10 second) as the sensor is working. The user sent information from an android mobile application; as a result, the LCD is displaying received data, as seen in Fig. 22b. These results demonstrate the efficiency of the proposed idea and give immediate validation for the proposed model. This is a prototype and these kinds of application can be implemented in the town, cities to let peoples alert from incidents.

We utilized Arduino-based automation systems with wireless communication, where street lights could be operated automatically based on solar rays, and wirelessly, etc. In some cases, we required more wireless connectivity to make the system more scalable for easy integration of new devices, and online access to the information managed by the Arduino boards are designed to build a user-friendly GUI. In this regard, the ethernet infrastructure would be used to reach other computers from the internet, and the ethernet module can allow a quick reach to these data. Meanwhile, to store and secure the acquired information of devices for further study and analysis, different kind of database is used in our proposed designs; in fact, Arduino is easy to integrate with many databases. Besides, the mentioned systems are viewed as lab-scale experiments, as noted above; thus, the regulated power for real-scale installations can be modified by driving an external voltage controller. Similarly, the proposed solutions/products can be operated for real-scale deployments by adding certain technologies such as BLE [114]. IR obstacle detection system is replaced by ultrasonic distance sensors with a maximum range of 4 m; they are inexpensive and sensitive enough for this form of application.

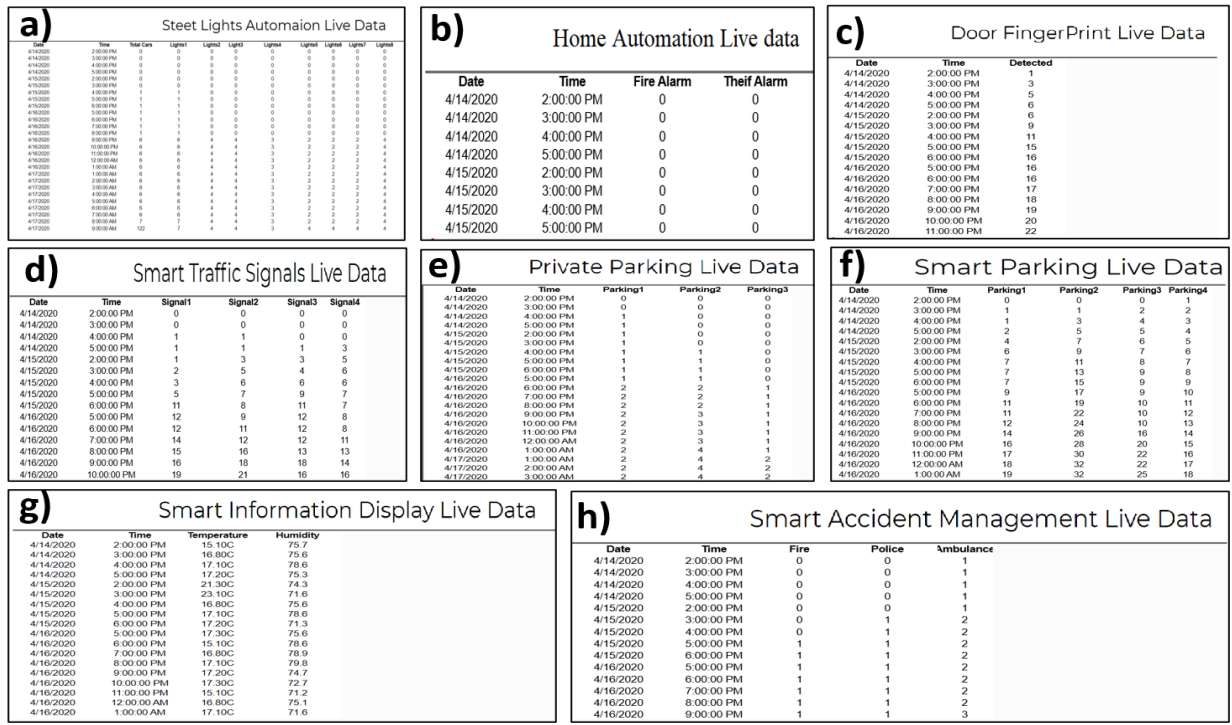

**FIGURE 23.** Data displaying on the website; (a) Street light automation data is illustrated; (b) Home automation data is displayed; (c) Door fingerprint data is demonstrated; (d) Traffic management system data; (e) Private parking data is shown; (f) Smart parking data is illustrated; (g) Data of the information display system is demonstrated; (h) Accident management data is shown.

It is worth remembering that objects going through the route also contain a wide variety of substances and are more difficult to track at night. Data is essential for today's world, as many problems and future prediction are quickly made by historical data. The historical record so that the user can evaluate various time intervals and patterns to make possible forecasts.

### 4.9. LIVE DATA FOR ALL SMART CITY SYSTEMS

However, to improve safety measurements, the serial monitor of Arduino IDE is further adopted and considered as an interface to count and display on the serial monitor and website such as the number of objects passing objects on the road, home appliances switching, and other operation as defined earlier. In this proposed system, the live data for all smart city systems are displayed in serial monitor and website, and also stored in excel files. For Demonstration purposes, Fig. 23 shows the total number of objects detected by all systems on the website (a locally stored website that can be easily hosted on the internet). Where in Fig. 23a, the street light automation data is displayed on the website page having ten attributes such as date, time and light 1–8. Home automation data is shown on the website having four attributes such as date, time, thief alarm and fire alarm, as illustrated in Fig. 23b. Door fingerprint data is shown with three attributes, as shown in Fig. 23c. In Fig. 23d, four-way road traffic signal data on the website is displayed. In Fig. 23e, Private parking data with three attributes is illustrated. In Fig. 23f, smart parking data with four attributes are shown. Information display data having temperature and humidity attributes are shown on the website, as displayed in Fig. 23g. In Fig. 23h, accident management having a fire, ambulance, and police attributes is display on the website is demonstrated. Table 2 summarizes the comparison of old systems and the proposed systems as a quick review.

Table 2. Comparison between old systems and the proposed system.

| Functionality | Old Systems | Proposed System |
|---|---|---|
| Based on Arduino | Yes | Yes |
| Streetlights operates with ultrasonic sensor and mobile application | No | Yes |
| Traffic signal skip if no traffic capability | No | Yes |
| Available parking slot display capability with vehicle's detection | No | Yes |
| Automatically sends Real time location of incident | No | Yes |
| Display and record all systems Live data in database and website. | No | Yes |

The proposed streetlight system has much affective in today's lives as the shortage of energy; it will be deployed in every city street lights, parking of shopping malls, housing societies. The effectiveness of the proposed android-based application would systematically introduce yet another aspect to human interactions with consumer electronic devices, such as fans, lights, cooler. The proposed mobile-application technique has much effective for different purposes, such as, in the medical field, it can be used by disabled patients to operate appliances without moving from place. Making an effective traffic control system led to the saving of time by individuals; they do not have to wait for a long time if other roads have no traffic. The effectiveness of the proposed smart and private parking system saves peoples time as well as money; they can easily find parking slots for their vehicles. The smart information display system has much use for the citizens of the city as they should be aware of daily incidents and news in the town, which is the solution to many problems. By using door fingerprint system, it saves much time and the trouble of carrying a key always with them; one should can and efficiently operate their doors with their fingerprints. The major effectiveness of accident management systems saves many lives, as the corresponding department will be aware of the incident in seconds, without any misunderstandings or wrong location as the system is entirely effective which also saves much time.

## 5. CONCLUSION

In conclusion, a design prototype / product for a low-cost multiple-applications system for a smart city based on Arduino has been demonstrated, which can be programmed to react to events (based on automatic mode and an android mobile application as described above) and to cause corresponding actions. The proposed system presented with mainly seven contributions in which the first system is prototype that uses an android application based controlling of street lights according to the mobile signals. Home automation system which is a product that can be directly installed in real-time based on android mobile application to control different appliances easily through the internet. Moreover, door fingerprint and home security products are also presented for safety measurements. After that, the four-way road traffic system is discussed that can skip or increase time interval based on the flow of traffic, and this technique can be directly installed in real-time. Private and smart system is products, which operated on the detection of vehicles and showed available parking slots could also be directly installed in real-time. Furthermore, the accident management prototype system is discussed, which is capable of sending the exact location of the incident to the corresponding department within seconds. Finally, smart information display system which is a prototype and shows the daily incidents within the city or town to let aware the citizen from hazards. The hardware implementation of the integrated systems was provided by a lab-scale model to show the usability, adaptability, durability, accuracy and low cost of the device. As a lesson to be learnt, we affirmed that the systems/products can easily be implemented on a wide scale under real conditions in the near future, and can be effectively deployed in cities, towns, malls, houses, housing communities, etc.

Meanwhile, the proposed multiple-application system has the advantages such as user-friendly, low-cost approach, Low power consumption, compact and fast to use, and the device is smaller in scale, so less space is needed to conform to the hardware circuits. Also, the designed prototype is highly robust against unexpected problems. It can be easily extended further in the hardware section, and various applications can be added to reduce the human effort of upgrading. Similarly, voice commands are sent and received through wireless serial transmission with the help of Bluetooth technology. On the other hand, as the range of Bluetooth technology is up to 10–15 m only, the distance of processing the proposed system is smaller. The delay in transmission and response of commands becomes high if the Bluetooth connection gets dropped frequently. The ultrasonic sensor has a range of 3 meters only, and it has glitch some time that they can detect air as a motion.

## 6. FUTURE WORK

Future work will build upon the improvement of the detection system to increase accuracy and more functionality. Ultrasonic senor can be replaced with long-range PIR motion detection sensor. A streetlight can also be controlled with Wi-Fi where long-distance matters. In addition, the power supply provided to the receiver can be provided through a solar power system as supply provided by solar power is in the form of direct current (DC) and receiver also works on DC. By using Wi-Fi, the transmitting range can be expanded when mounting routers over short distances and using a GSM kit for wireless transmission. An on-board wireless camera can be mounted and provides live streaming and can be used to monitor the car from remote locations. The camera can be deployed at traffic signals, parking and accident management system to make more efficient and image processing and pattern classifying techniques can be used to better accuracy to detection.

## FUNDING

This work was supported by Khwaja Fareed University of Engineering & Information Technology, Rahim Yar Khan, Pakistan.

## ACKNOWLEDGMENT

The authors would like to thank the Editor and the anonymous reviewers for their insightful comments and constructive suggestions that certainly improved the quality of this paper. The authors thank Zaid Mumtaz from Department of Computer Science, Superior University, Lahore, 55150, Pakistan and Majid Saleem, from Department of Computer Science, Khwaja Fareed University of Engineering and Information Technology, Rahim Yar Khan 64200, Pakistan for critical discussions.

**Conflict of interest:** Authors declear no conflict of interest.

## REFERENCES

[1] K. Axelsson, and M. Granath, "Stakeholders' stake and relation to smartness in smart city development: Insights from a Swedish city planning project," *Gov. Inf. Q.*, vol. 35, no. 4, pp. 693–702, 2018, Doi: 10.1016/j.giq.2018.09.001.

[2] H. Arasteh, V. Hosseinnezhad, V. Loia, A. Tommasetti, O. Troisi, M. Shafie-khah, and P. Siano, "Iot-based smart cities: A survey," 16th *Int. Conf. Environ. Electr. Eng.,* Florence, 2016, pp. 1-6.

[3] N. C. Luong, D. T. Hoang, P. Wang, D. Niyato, D. I. Kim, and Z. Han, "Data Collection and Wireless Communication in Internet of Things (IoT) Using Economic Analysis and Pricing Models: A Survey," *IEEE Commun. Surv. Tutorials*, vol. 18, no. 4, pp. 2546–2590, 2016, Doi: 10.1109/COMST.2016.2582841.

[4] W. M. Da Silva, G. H. R. P. Tomas, K. L. Dias, A. Alvaro, R. A. Afonso, and V. C. Garcia, "Smart cities software architectures: A survey," in *Proc. ACM Symp. Appl. Comput.*, 2013, pp. 1722–1727, Doi: 10.1145/2480362.2480688.

[5] S. Pellicer, G. Santa, A. L. Bleda, R. Maestre, A. J. Jara, and A. G. Skarmeta, "A global perspective of smart cities: A survey," *7th Int. Conf. Innov. Mob. Internet Serv. Ubiquitous Comput.*, 2013, pp. 439–444, Doi: 10.1109/IMIS.2013.79.

[6] R. Petrolo, V. Loscrì, and N. Mitton, "Towards a smart city based on cloud of things, a survey on the smart city vision and paradigms," *Trans. Emerg. Telecommun. Technol.*, vol. 28, no. 1, 2017, Doi: 10.1002/ett.2931.

[7] Giordano, G. Spezzano, and A. Vinci, "Smart agents and fog computing for smart city applications," *Proc. First Int. Conf. Smart Cities,* Jun. 2016, pp. 13–127, Doi: 10.1007/978-3-319-39595-1.

[8] L. Wang, and D. Sng, "Deep learning algorithms with applications to video", 10-Dec-2015. [Online]. Available: https://arxiv.org/pdf/1512.03131.pdf. [Accessed: 22-Apr-2020].

[9] F. S. Sutil, and A. Cano-Ortega, "Smart public lighting control and measurement system using lora network," *Electron.*, vol. 9, no. 124, 2020, Doi: 10.3390/electronics9010124.

[10] Z. Mumtaz, S. Ullah, Z. Ilyas, N. Aslam, S. Iqbal, S. Liu, j. A. Meo, and H. A. Madni, "An automation system for controlling streetlights and monitoring objects using arduino," *Sens.*, vol. 18, no. 178, pp. 1–14, 2018, Doi: 10.3390/s18103178.

[11] L. B. Imran, R. M. A. Latif, M. Farhan, and T. Tariq, "Real-time simulation of smart lighting system in smart city," Int. J. Space-Based Situated Comput., vol. 9, no. 2, pp. 90, 2019, Doi: 10.1504/ijssc.2019.104219.

[12] M. Beccali, M. Bonomolo, G. Ciulla, A. Galatioto, and V. Lo Brano, "Improvement of energy efficiency and quality of street lighting in South Italy as an action of sustainable energy action plans. The case study of Comiso (RG)," Energy, vol. 92, pp. 1-15, 2015, Doi: 10.1016/j.energy.2015.05.003.

[13] S. R. Parekar, and M. M. Dongre, "An intelligent system for monitoring and controlling of street light using GSM technology," IEEE Int. Conf. Inf. Process, Dec. 2015, pp. 604–609.

[14] X. Liang, S. Li, and J. Fei, "Adaptive fuzzy global fast terminal sliding mode control for microgyroscope system," IEEE Access, vol. 4, pp. 9681–9688, 2016, Doi: 10.1109/ACCESS.2016.2636901.

[15] B. Abinaya, S. Gurupriya, and M. Pooja, "Iot based smart and adaptive lighting in street lights," Int. Conf. Comput. Commun. Technol., 2017, pp. 195–198, Doi: 10.1109/ICCCT2.2017.7972267.

[16] S. patel, Ruderesh, Kallendrachari, M. K. Kumar, and Vani, "Design and implementation of automatic street light control using sensors and solar panel," Int. J. Eng. Res. Appl., vol. 5, no. 6, pp. 97-100, Jun. 2015.

[17] M. Wadi, A. Shobole, M. R. Tur, and M. Baysal, "Smart hybrid wind-solar street lighting system fuzzy based approach," Int. Istanbul Smart Grids Cities Congress Fair, Istanbul, Turkey , 2018, pp. 71–75.

[18] Components101 "LDR", 26-Sep.-2017. [Online]. Available: https://components101.com/5v-relay-pinout-working-datasheet [Accessed: 24-Apr.-2020].

[19] components101 "IR Sensor Module", 30-Aug.-2018. [Online]. Available: https://components101.com/sensors/ir-sensor-module [Accessed: 24-Apr.-2020].

[20] L. Louis, "Working principle of arduino and using it as a tool for study and research", Int. J. Control Autom. Commun. Syst., vol. 1, pp. 21–29, 2016.

[21] Arduino "Arduino Uno", 2020. [Online]. Available: https://store.arduino.cc/usa/arduino-uno-rev3/ [Accessed: 24-Apr.-2020].

[22] R. Salvi, S. Margaj, K. Mate, and P.B. Aher, "Smart street light using arduino uno microcontroller", Int. J. Innov. Res. Comput. Commun. Eng., vol. 5, pp. 5203–5206, 2017.

[23] P.C. Cynthia, V.A. Raj, S.T. George, "Automatic street light control based on vehicle detection using arduino for power saving applications", Int. J. Electron. Electr. Comput. Syst., vol. 6, pp. 291–295, 2017.

[24] C. Thapa, D. Rasaily, Wangchuk, T.R. Wangchuk, A. Pradhan, and A. Ashraf, "Auto intensity control of street light with solar tracker using microcontroller," Int. J. Eng. Trends Technol. vol.33, pp. 369–372, 2016.

[25] R. Banerjee, "Solar tracking system," Int. J. Sci. Res. Publ., vol. 4, pp. 1-7, 2015.

[26] T. Rajasekhar, K.P. Rao, "Solar powered led street light with auto intensity control," Int. J. Tech. Innov. Mod. Eng. Sci., vol. 3, pp.1-4, 2017.

[27] M. Srikanth, and K.N. Sudhakar, "Zigbee based remote control automatic street light system," Int. J. Eng. Sci. Comput., pp. 639–643, 2014.

[28] A. Rao, and A. Konnur, "Street light automation system using arduino uno," Int. J. Innov. Res. Comput. Commun. Eng., vol. 5, pp. 16499–16507, 2017.

[29] Sravani, P. Malarvezhi, and R. Dayana, "Design and implementation of dimmer based smart street lighting system using raspberry pi and IoT," Int. J. Eng. Technol., vol. 7, pp. 524-528, 2018.

[30] T. D. P. Mendes, R. Godina, E. M. G. Rodrigues, J. C. O. Matias, and J. P. S. Catalão, "Smart home communication technologies and applications: Wireless protocol assessment for home area network resources." Energies, vol.8, pp. 7279-7311, July 2015, Doi: 10.3390/en8077279

[31] V. Sundaram, R. M., P. M., and V. S. J., "Encryption and hash based security in Internet of Things," 3rd Int. Conf. Signal Process. , Commun. Networking (ICSCN), 2015, pp. 1–6 .

[32] S. SMIEEE, and Veena M, "Implementation of interactive home automation system based on e-mail and bluetooth technologies", Int. Conf. Inf. Process, 2015, pp. 458–463.

[33] K. Gill, S. Yang, F. Yao, and X. Lu, "A ZigBee-based home automation system," IEEE Trans. Consum. Electron., vol.55 , no. 2, May 2009, Doi: 10.1109/TCE.2009.5174403.

[34] P. N. Adhagale, "Smart home automation system using ethernet technology," Int. J. Innovative Res. Sci. Eng. Technol. , vol.6 , no.11, pp. 21717–21722, Nov. 2017, Doi: 10.15680/IJIRSET.2017.0611110.

[35] N. K. Kaphungkui, "RF based Remote Control for Home Electrical Appliances," Int. J. Innovative Res. Electr. , Electron. Instrum. Control Eng. , vol. 3, no. 7, pp. 1–4, Jul. 2015 , Doi:10.17148/IJIREEICE.2015.3709.

[36] G. J. Rao, A. Vinod, N. Priyanka, C. Siva, and H. Kumar, "IOT based web controlled home automation using raspberry PI," Int. J. Sci. Res. Sci, Eng. Technol., vol. 6 ,no. 2, pp. 229-234 , Mar. 2019, Doi: 10.32628/IJSRSET196246.

[37] Y. Mittal , P. Toshniwal , S. Sharma, D. Singhal, R. Gupta, and V. K. Mittal, "A voice-controlled multi-functional smart home automation system," Annu. IEEE India Conf. (INDICON), New Delhi, 2015, pp. 1-6.

[38] A. Kassem, S. El Murr, G. Jamous, E. Saad, and M. Geagea, "A Smart Lock System using Wi-Fi Security," 3rd Int. Conf. Adv. Comput. Tools Eng. Appl., Beirut, 2016, pp. 222-225.

[39] L. Kamelia, A. N. S. R, M. Sanjaya, and E. Mulyana, "Door-automation system using bluetooth-based android for mobile phone," ARPN J. Eng. Appl. Sci. , vol. 9, no. 10, pp. 1759–1762, Oct. 2014.

[40] U. Khan, M. Arif, M. F. Khalid, R. Ali, Q. Khan, and S. Ahmad, "Voice controlled home automation system," Int. J. Res. comput. Commun. Technol , vol.6 , no. 5 , May 2017.

[41] Nandeesh , B. S. Reddy and S. Kumar , "Intelligent security system for industries by using GPS and GSM," Int. J. Adv. Res. Comput. Sci. Technol. , vol. 2, no. 1, pp. 3–5, 2014.

[42] S. MT, N. Ilahi, B. Musrawati, Syarifuddin, A. Achmad, and E. Umrianah, "Early Leakage Protection System of LPG ( Liquefied Petroleum Gas ) Based on ATMega 16 Microcontroller," IOP Conf. Ser. Mater. Sci. Eng. , vo. 336, 2018, pp. 1-11.

[43] M. Sathishkumar and S. Rajini, "Smart surveillance system using PIR sensor network and GSM," Int. J. Adv. Res. Comput. Eng. Technol. , vol. 4, no. 1, 2015.

[44] B. Pandurang, J. Dhanesh, P. M. S. Pede, G. Akshay, and G. Rahul, "Smart lock : A locking system using bluetooth technology & camera verification," Int. J. Tech. Res. Appl., vol. 4, no. 1, pp. 136–139, 2016.

[45] S. Djahel, R. Doolan, G. M. Muntean, and J. Murphy, "A communications-oriented perspective on traffic management systems for smart cities: Challenges and innovative approaches," IEEE Commun. Surv. Tutorials, vol. 17, no. 1, pp. 125–151, 2015.

[46] Al Nuaimi, H. Al Neyadi, N. Mohamed, and J. Al-jaroodi, "Applications of big data to smart cities," J. Internet Serv. Appl., pp. 1-15 , Mar. 2016, Doi: 10.1186/s13174-015-0041-5.

[47] B. Hammi, R. Khatoun, S. Zeadally, A. Fayad, and L. Khoukhi, "IoT technologies for smart cities," Inst. Eng. Technol., vol. 7, no.1, pp. 1–13, 2018, Doi: 10.1049/iet-net.2017.0163.

[48] S. S. Mahalingam and S. Arockiaraj, "Density based traffic light control using arduino," Int. J. Adv. Res. Innovative Ideas Educ., vol.4, no. 5, pp. 805–812, 2018.

[49] Y. Shinde and H. Powar, "Intelligent traffic light controller using IR sensors for vehicle detection," Int. Adv. Res. J. Sci. Eng. Technol., vol.4, no.2, pp. 2393–2395, Jan. 2017, Doi: 10.17148/IARJSET/NCETETE.2017.28.

[50] M. Ehsan, "Smart traffic light controller based on microcontroller," Iranian J. Chem. Chem. Eng., vol.16, no.1 , Jan. 2016.

[51] L. Y. Deng, N. C. Tang, D. L. Lee, C. T. Wang, and M. C. Lu, "Vision based adaptive traffic signal control system development," Int. Conf. Adv. Inf. Networking Appl., 2005, pp. 5–8.

[52] M. Vidhyia, S. Elayaraja, M. Anitha, M. Divya, and S. D. Barathi, "Traffic light control system using raspberry-PI," Asian J. Electr. Sci., vol. 5, no. 1, pp. 8–12, 2016.

[53] O. Emeand, C. E. Madubuike, J. O. Idemudia, A. U. Ntuen, "Simulation of N-way traffic lights ising arduino uno environment," Int. J. Comput. Appl. Technol. Res., vol. 5, no. 8, pp. 543–550, 2016.

[54] M. Bogdan, "Traffic light using arduino uno and LabVIEW," 12th Int. Conf. Virtual Learn, 2017 , pp. 286-290 .

[55] S. M. Kadam and T. Engineering, "Raspberry Pi based signal breaking and e-mail system," Int. J. Adv. Res. Innovative Ideas Educ., vol. 4, no. 2, pp. 52–57, 2018.

[56] Basil and P. S. D. Sawant, "IoT based traffic light control system using Raspberry Pi," Int. Conf. Energy Commun. Data Anal. Soft Comput, 2017, pp. 1078–1081.

[57] A. Agrawal, and S. Saurabh, "Traffic control system using Zigbee module," Int. J. Eng. Sci. Comput., vol. 7, pp. 11150-11153, 2017.

[58] P. A. Saiba, M. U. Afeefa, T. S. Aruna, C. Jose, and V. M. Radhika, "Density based traffic signal system using PIC microcontroller," Int. J. Comput. Trends Technol., vol. 47, no. 1, pp. 74–78, 2017.

[59] A. Chattaraj, S. Bansal and A. Chandra, "An intelligent traffic control system using RFID," IEEE Potentials, vol. 28, no. 3, pp. 40-43, May-June 2009.

[60] M. I. Bakthansari, "Traffic control system for two way lane based ON GSM module," Int. J. Sci. Eng. Res., vol. 8, no. 4, pp. 1054-1057, 2017.

[61] R. Jamal, K. Manaa, M. Rabee' a, and L. Khalaf, "Traffic control by digital imaging cameras," Emerging Trends Image Process., Comput. Vision Pattern Recognit., pp. 231-247, 12-Dec.-2014. [Online]. Available: http://dx.doi.org/10.1016/B978-0-12-802045-6.00015-6.

[62] Alegria and P. S. Girão, "Vehicle plate recognition for wireless traffic control and law enforcement system," IEEE Int. Conf. Ind. Technol., 2016 , pp. 1800–1804.

[63] A. Bin Sulaiman, M. F. B. M. Afif, M. A. Bin Othman, M. H. Bin Misran, and M. A. B. M. Said, "Wireless based smart parking system using zigbee," Int. J. Eng. Technol., vol. 5, no. 4, pp. 3282–3300, 2013.

[64] Marso and D. Macko, "A new parking-space detection system using prototyping devices and bluetooth low energy communication," Int. J. Eng. Technol. Innov., vol. 9, no. 2, pp. 108–118, 2019.

[65] R. Salpietro, L. Bedogni, M. Di Felice, and L. Bononi, "Park Here! A smart parking system based on smartphones' embedded sensors and

short range Communication Technologies", IEEE World Forum Internet Things, pp. 18–23, 2015, Doi: 10.1109/WF-IoT.2015.7389020.
[66] S. Nandyal, S. Sultana, and S. Anjum, "Smart car parking system using arduino UNO," Int. J. Comput. Appl., vol. 169, no. 1, pp. 13–18, 2017, Doi: 10.5120/ijca2017914425.
[67] A. AlHarbi, B. AlOtaibi, M. Baatya, Z. Jastania, and M. Meccawy, "A smart parking solution for Jeddah city," Int. J. Comput. Appl., vol. 171, no. 7, pp. 4–9, 2017, Doi: 10.5120/ijca2017915084.
[68] Gupta, P. Rastogi, and S. Jain, "Smart parking system using cloud based computation and raspberry Pi, 2nd Int. Conf. I-SMAC (IoT Soc. Mobile, Anal. Cloud), I-SMAC India, 2018, pp. 94-99, Doi: 10.1109/I-SMAC.2018.8653764.
[69] Sucharitanjani, P. N. Kumar, and Bhupathi, "Internet of things based smart vehicle parking access system," Int. J. Innov. Technol. Explor. Eng., vol. 8, no. 6, pp. 732–734, 2019.
[70] P. A. Gavali, P. Kunnure, S. Jadhav, T. Tate, and V. Patil, "Smart Parking System Using the Raspberry Pi and Android," Int. J. Comput. Sci. Inf. Technol. Res., vol. 5, no. 2, pp. 48–52, 2017.
[71] P. S. Reddy, G. S. N. Kumar, B. Ritish, C. Saiswetha, and K. B. Abhilash, "Intelligent Parking Space Detection System Based on Image Segmentation," Int. J. Sci. Res. Dev., vol. 1, no. 6, pp. 1310-1312, 2013.
[72] M. M. Rashid, A. Musa, M. A. Rahman, N. Farahana, and A. Farhana, "Automatic parking management system and parking fee collection based on number plate recognition," Int. J. Mach. Learn Comput., vol. 2, no. 2, Apr. 2012.
[73] P. S. S. Thorat, M. Ashwini, A. Kelshikar, and S. Londhe, "IoT based smart parking system using RFID," Int. J. Comput. Eng. Res. Trends, vol. 4, no. 1, pp. 9–12, 2017, Doi: 10.22362/ijcert/2017/v4/i1/xxxx.
[74] Y. Rahayu, and F. N. Mustapa, "A secure parking reservation system using GSM technology," Int. J. Comput. Commun. Eng., vol. 2, no. 4, pp. 2–5, Jul. 2013, Doi: 10.7763/IJCCE.2013.V2.239.
[75] M. H. Kadhim, "Arduino smart parking manage system based on ultrasonic internet of things (IoT ) technologies," Int. J. Eng.Technol., vol. 7, pp. 494–501, 2018.
[76] R. F. O. Salma, and M. M. Arman, "Smart parking guidance system using 360° camera and Haar-Cascade classifier on IoT system," Int. J. Recent Technol Eng., vol. 8 , no. 2, pp. 864–872, Sep. 2019, Doi: 10.35940/ijrte.B1142.0982S1119.
[77] Kaur, "Implementation of smart parking using artificial intelligence," Int. J. Sci. Dev. Res., vol. 4, no. 8, pp. 284–290, 2019.
[78] S. Aarya, C. K. Athulya, P. Anas, B. Kuriakose, J. S. Joy, and L. Thomas, "Accident alert and tracking using arduino," Int. J. Adv. Res. Electr. Electron. Instrum. Eng., vol. 7, no. 4, pp. 1671–1674, Apr. 2018, Doi: 10.15662/IJAREEIE.2018.0704018.
[79] S. Janakiramkumaar, A. Manavalan, S. Rajarajan, and R. Premalatha, "Accident detection and vehicle safety using Zigbee," Int. Res. J. Eng. Technol., vol. 3, no. 5, pp. 947–950, 2018.
[80] KumarA, U. Scholar, and A. Professor, "Accident Detection and Alerting System Using Gps & Gsm," Int. J. pure Appl. Math., vol. 119, no. 15, pp. 885–891, 2018.
[81] S. Sharma, and S. Sebastian, "IoT based car accident detection and notification algorithm for general road accidents," Int. J. Electr. Comput. Eng., vol. 9, no. 5, pp. 4020–4026, 2019, Doi: 10.11591/ijece.v9i5.pp4020-4026.
[82] A. Tsuge, H. Takigawa, H. Osuga, H. Soma, and K. Morisaki, "Accident vehicle automatic detection system by image processing technology," Proc. VNIS'94 Veh. Navig. Inf. Syst. Conf., Yokohama, Japan, 1994, pp. 45-50.
[83] Y. Ki, "Accident detection system using image processing and MDR," Int. J. Comput. Sci. Netw. Secur., vol. 7, no. 3, pp. 35–39, Mar. 2007.
[84] Dogru, and A. Subasi, "Traffic accident detection using random forest classifier," Learn. Technol. Conf., 2018, pp. 40–45.
[85] Thompson, J. White, B. Dougherty, A. Albright, and D. C. Schmidt, "Using smartphones and wireless mobile sensor networks to detect car accidents and provide situational awareness to emergency responders," Lect. Notes Inst. Comput. Sci. Social Inf. Telecommun. Eng. Mobile Wireless Middleware Operating Syst. Appl., 2010, pp. 29–42.
[86] Zaldivar, C. T. Calafate, J. C. Cano, and P. Manzoni, "Providing accident detection in vehicular networks through OBD-II devices and android-based smartphones," Conf. Local Comput. Networks, LCN, 2011, pp. 813–819, Doi: 10.1109/LCN.2011.6115556.
[87] R. R. Reddy, N. Prashanth, G. Indira, and M. Sharada, "Electronic Scrolling Display Using Arduino Board," Int. J. Eng. Res. Electr. Electron. Eng., vol. 4, no. 2, pp. 49–52,  Feb. 2018.
[88] T. Prakash, K. N. Ayaz, and O. P. Sumtilal, "Digital notice board ," Int. J.  Eng. Dev. Res., vol. 5, no. 2, pp. 127–130, 2017.
[89] D. Chakraborty, S. Yadav, S. Rathore, S. Kumar, R. Agrawal, and P. Chandrakar "Smart rolling LED display using arduino and bluetooth," Int. J. Sci. Res. Comput. Sci. Eng. Inf. Technol., vol. 2, no. 3, pp. 322–324, 2017.
[90] A. Yedilkhan, S. Rassim, B. Andrey, and N. Ari, "Design of an information display based on several LED matrices and a single microcontroller," Int. Conf. Appl. Inf. Commun. Technol, Baku, 2013, pp. 1-4.
[91] E. Alam, M. A. Kader, S. A. Proma, and S. Sharma, "Development of a Voice and SMS Controlled Dot Matrix Display Based Smart Noticing System with RF Transceiver and GSM Modem," 21st Int. Conf. Comput. Inf. Technol., Dhaka, Bangladesh, 2018, pp. 1-5.
[92] Tanadumrongpattana,  A.  Suethakorn, S.  Mitatha,  and  C. Vongchumyen, "SMS information display board," Int. Sci. Social Sci Eng. Energy Conf., 2011, pp. 186-189.
[93] P. C. N. Bhoyar, S. Khobragade, and S. Neware "Zigbee Based Electronic Notice Board," Int. J. Eng. Sci. Comput., vol. 7, no. 3, pp. 5422–5424, Mar. 2017.
[94] P. P. Kulkarni, S.V.Patil, B. R. Shingate, V. N. Mali, S. S.Thoke, S. N. Pawar, "Wireless Digital Electronic Notice Board Using Wi-Fi," Int. J. Innov. Eng. Res. Technol., vol. 5, no. 4, pp. 47–52, 2018.
[95] A. Anish, A. A. Merlin, G. J. J. Malar, and J. Jeseeba, "Wireless notice board using arduino and IOT technology," Int. Conf. Recent Trends Electr. Electron. Eng., 2019, pp. 99–105, Doi: 10.23883/IJRTER.CONF.20190304.016.O5THY.
[96] M. Zungeru, G. D. Obikoya, O. F. Uche, and T. Eli, "Design and implementation o f a GSM-Based scrolling message display board," Int. J. Comput. Sci. Inf. Technol. Control Eng., vol. 1, no. 3, pp. 21–31, 2014.
[97] Al Dahoud, Ali & Fezari, Mohamed. (2018). NodeMCU V3 For Fast IoT Application Development.
[98] Sparkfun "specification of lcd module", 29-10.-2008. [Online]. Available: https://www.sparkfun.com/datasheets/LCD/ADM1602K-NSW-FBS-3.3v.pdf [Accessed: 24-Apr.-2020].
[99] components101 "DHT11–Temperature and Humidity Sensor", 5-Jan.-2018. [Online]. Available: https://components101.com/dht11-temperature-sensor [Accessed: 24-Apr.-2020].
[100] Nxp  "MFRC522", 27-April-2016. [Online]. Available: https://www.nxp.com/docs/en/data-sheet/MFRC522.pdfsensor [Accessed: 24-Apr.-2020].
[101] Itead "Serial Port Bluetooth Module HC-05",  24-March-2017. [Online].  Available: https://www.itead.cc/wiki/Serial_Port_ Bluetooth_Module_(Master/Slave)_:_HC-05   [Accessed: 24-Apr.-2020].
[102] components101 "Servo Motor SG-90", 18-Sep.-2018. [Online]. Available: https://components101.com/servo-motor-basics-pinout-datasheet [Accessed: 24-Apr.-2020].
[103] components101 "HC-SR04 Ultrasonic Sensor", 18-Sep.-2018. [Online]. Available: https://components101.com/ultrasonic-sensor-working-pinout-datasheet [Accessed: 24-Apr.-2020].
[104] Simcom "Sim800l hardware design v1.00", 20-Aug.-2013. [Online]. Available: https://img.filipeflop.com/files/download/Datasheet_SIM800L.pdf [Accessed: 24-Apr.-2020].
[105] components101 "NEO-6MV2 GPS Module", 16-Oct.-2018. [Online]. Available: https://components101.com/modules/neo-6mv2-gps-module [Accessed: 24-Apr.-2020].
[106] Adafruit learning system "Adafruit Optical Fingerprint Sensor", 28-Oct.-2019. [Online]. Available: https://cdn-learn.adafruit.com/downloads/pdf/adafruit-optical-fingerprint-sensor.pdf [Accessed: 24-Apr.-2020].
[107] components101 "LDR", 30-Oct.-2017. [Online]. Available: https://components101.com/ldr-datasheet [Accessed: 24-Apr.-2020].

[108] components101 "5V 5-Pin Relay", 26-Sep.-2017. [Online]. Available: https://components101.com/5v-relay-pinout-working-datasheet [Accessed: 24-Apr.-2020].
[109] Google Drive "Steetlights", 24-Apr.-2020. [Online]. Available: https://drive.google.com/open?id=1J-ndQaC7RyrzwweZOvysDMBzWNhG3f19 [Accessed: 24-Apr.-2020].
[110] Google Drive "Blynk", 24-Apr.-2020. [Online]. Available: https://drive.google.com/open?id=1Ih9MvKbuaf-I3_jOKcoGmrCwyamqND6Z [Accessed: 24-Apr.-2020].
[111] Google Drive "Smart Info", 24-Apr.-2020. [Online]. Available: https://drive.google.com/open?id=1hElFtcjNAz2mKuLIbVo9E8zuMsBubwTW [Accessed: 24-Apr.-2020].
[112] S.C. Pokress, and J.J.D. Veiga, "MIT app inventor enabling personal mobile computing," 2013. [Online]. Available: https://arxiv.org/abs/1310.2830 [Accessed: 24-Apr.-2020].
[113] Youtube " Design and Implementation of Low-Cost Smart City Using Microcontrollers", 4-Oct.-2020. [Online]. Available: https://www.youtube.com/watch?v=evviwwy-03s&feature=youtu.be
[114] K. Mikhaylov, N. Plevritakis, and J. Tervonen, "Performance analysis and comparison of bluetooth low energy with ieee 802.15.4 and simplicity," *J. Sens. Actuator Netw.,* vol. 2, pp. 589–613, 2013.